\documentclass{aa}
\usepackage{graphicx,longtable,lscape}
\usepackage{txfonts,psfig}
\usepackage[]{natbib}
\def \sw {{\it Swift}}

\def \ferg {erg cm$^{-2}$ s$^{-1}$}
\def \hcm {\hbox {\ifmmode $ atom cm$^{-2}\else atom cm$^{-2}$\fi}}

\begin{document}
   \title{The Palermo {\it Swift}-BAT hard X-ray catalogue\\} 
   \subtitle{III. Results after 54 months of sky survey\\}
   \author{G.\ Cusumano\inst{1}, V.\ La Parola\inst{1}, 
           A.\ Segreto\inst{1},  C.\ Ferrigno\inst{1,2,3}, A.\ Maselli\inst{1},
	   B.\ Sbarufatti\inst{1}, P.\ Romano\inst{1}, 
	   G. Chincarini\inst{4,5}, P. Giommi\inst{6}, N. Masetti\inst{7}, A. Moretti\inst{5},
	   P. Parisi\inst{7,8}, G. Tagliaferri\inst{5}}

   \offprints{G. Cusumano, cusumano@ifc.inaf.it}
   \institute{INAF, Istituto di Astrofisica Spaziale e Fisica Cosmica di Palermo, 
	Via U.\ La Malfa 153, I-90146 Palermo, Italy 
      \and
          Institut f\"ur Astronomie und Astrophysik T\"ubingen (IAAT)
	\and
          ISDC Data Centre for Astrophysics,
          Chemin d'\'Ecogia 16,
          CH-1290 Versoix,
          Switzerland
      \and
	  Universit\`a degli studi di Milano-Bicocca, Dipartimento di Fisica, Piazza delle Scienze 3, I-20126 Milan, Italy
      \and
	  INAF -- Osservatorio Astronomico di Brera, Via Bianchi 46, 23807 Merate, Italy
      \and
	  ASI Science Data Center, via Galileo Galilei, 00044 Frascati, Italy
      \and
          INAF, Istituto di Astrofisica Spaziale e Fisica Cosmica di Bologna, 
          via Gobetti 101, I-40129 Bologna, Italy
      \and
          Dipartimento di Astronomia, Universit\`a di Bologna,
          Via Ranzani 1, I-40127 Bologna, Italy 
     	  }
            
   \date{}
\abstract
{}
{We present the Second Palermo {\it Swift}-BAT hard X-ray catalogue obtained by
analysing data acquired in the first 54 months of the \sw\ mission.
}
{Using our software dedicated to the analysis of data from coded mask telescopes, 
we analysed the BAT survey data in three energy bands (15--30 keV, 15--70 keV, 15--150 keV), 
obtaining a list of 1256 detections above a significance threshold of 4.8 standard deviations. 
The identification of the source counterparts is pursued using two strategies:
the analysis of field observations of soft X-ray instruments and cross-correlation
of our catalogue with source databases.
}
{The survey covers 50\% of the sky to a 15--150 keV 
flux limit of  $1.0\times10^{-11}$ \ferg\ and $9.2\times10^{-12}$ 
\ferg\ for $|{\rm b}|< 10^{\circ}$  and $|{\rm b}|> 10^{\circ}$, respectively.
The  Second Palermo {\it Swift}-BAT hard X-ray catalogue includes 1079 ($\sim86$\%) hard X-ray sources 
with an associated counterpart (26 with a double association and 2 with a triple association) and 
177 BAT excesses ($\sim14$\%) that still lack a counterpart. 
The distribution of the BAT sources among the different object classes consists of $\sim19\%$
Galactic sources, $\sim57\%$ extragalactic sources, and
$\sim10\%$ sources with a counterpart at softer energies whose nature has not yet been
determined.
About half of the  BAT associated sources lack a counterpart in the ROSAT catalogues. 
This suggests that either  moderate or strong absorption may be preventing their detection in the 
ROSAT energy band. 
The comparison of our BAT catalogue with the Fermi Large Area Telescope First Source
Catalogue identifies 59 BAT/Fermi correspondences: 48 blazars,  3 Seyfert galaxies, 1
interacting galaxy, 3 high mass X-ray binaries, and 4 pulsars/supernova remnants. 
This small number of correspondences indicates that different populations make the sky shine
in these two different energy bands.}
{}
\authorrunning {G.\ Cusumano et al.}
\titlerunning {The 54-month Palermo \sw-BAT Hard X-ray Catalogue}

\maketitle

\section{Introduction\label{intro} }

The Burst Alert Telescope (BAT; \citealp{bat}) onboard the {\it Swift} observatory \citep{swift}
is a coded-aperture imaging camera operating in the 15--150 keV energy range  with a
large field of view (1.4 steradian half coded)  and a point spread function
(PSF) of 17 arcmin (full width half maximum). The telescope is mainly devoted to the monitoring 
of a large fraction of the sky for the occurrence of gamma ray bursts (GRBs). 
While waiting for new GRBs, BAT continuously collects spectral and
imaging information about the sky, covering a fraction of between 50\% and 80\% of the sky every 
day, providing the opportunity for a substantial gain in our knowledge of the Galactic and 
extragalactic sky  in the hard X--ray domain and increasing the sample of objects that
contribute to the luminosity in this energy range.
The first results of the BAT survey were presented in
\citet{markwardt05}, \citet{ajello1}, \citet{ajello3}, \citet{tueller08}, \citet{tueller09},
\citet{cusumano09}, and \citet{maselli10}. 
The First Palermo {\it Swift}-BAT hard X-ray catalogue \citep{cusumano09}
contains a list of 754 hard X-ray sources with an associated counterpart detected in the 
first 39 months of the {\it Swift} mission.
Among them, $\sim69$\% are extragalactic, $\sim27$\% are Galactic objects, $\sim4$\% are 
already known X-ray or $\gamma$-ray emitters whose nature has not yet been determined.

In this paper, we provide the Second Palermo {\it Swift}-BAT hard X-ray catalogue obtained
from the analysis of the data relative to the first 54 months of the {\it Swift} mission and
including 1256 BAT high-energy sources. The paper is organised as follows: 
in Sect.~\ref{datascreen}, we describe the screening and the analysis of the BAT survey data
and the global survey properties; in Sect.~\ref{detstrategy}, we illustrate our analysis strategy; 
in Sect.~\ref{id}, we describe the counterpart association strategy; the 54-month
catalogue and its properties are described in Sects.~\ref{cat} and \ref{catprop}.
Then, in Sect.~\ref{concl} we summarise our results.

The cosmology adopted in this work assumes $H_0=70$ km s$^{-1}$ Mpc$^{-1}$,
k=0, $\Omega_m=0.3$, and $\Lambda_0=0.7$. Quoted errors are at $1\sigma$
confidence level, unless otherwise specified.

\section{The BAT survey data\label{datascreen}}

The results presented in this paper were obtained by analising
the first 54 months of BAT survey data,
from 2004 December to the end of 2009 May. 
The data were retrieved from the {\it Swift} public 
archive\footnote{http://heasarc.gsfc.nasa.gov/cgi-bin/W3Browse/swift.pl} 
in the form of detector plane histograms 
(DPH): three-dimensional arrays (two spatial dimensions, 
one spectral dimension) that collect count-rate data in 
5-minute time bins for 80 energy channels.

To process the survey data, we developed and applied a code 
that performs screening, mosaicking, and source detection on data from coded mask instruments. 
This code is described in detail in \citet{segreto10}.
To screen out poor quality files from the data set, we rejected the DPHs:
\begin{itemize}
\item  with unstable spacecraft attitude (i.e., with a significant variation in
pointing coordinates).

\item produced near the SAA and characterized by a count rate much higher than the average value.

\item affected by inaccurate position reconstruction.
This was verified through a pre-analysis procedure where the position of the sources detected in single DPHs
was checked against a list of hard X-ray sources and transients (see \citealp{cusumano09} for details).
\item that were very noisy, i.e., with a standard deviation 
in the count rate (subtracted of both bright sources and background)  
significantly larger (a factor of 2) than that expected from statistics.
\end{itemize}

After the screening based on these criteria, the usable archive has a total
nominal exposure time of $\sim$100 Ms, corresponding to $\sim92$\% of the total survey
exposure time during the period under investigation.


\begin{figure}
\centerline{\psfig{figure=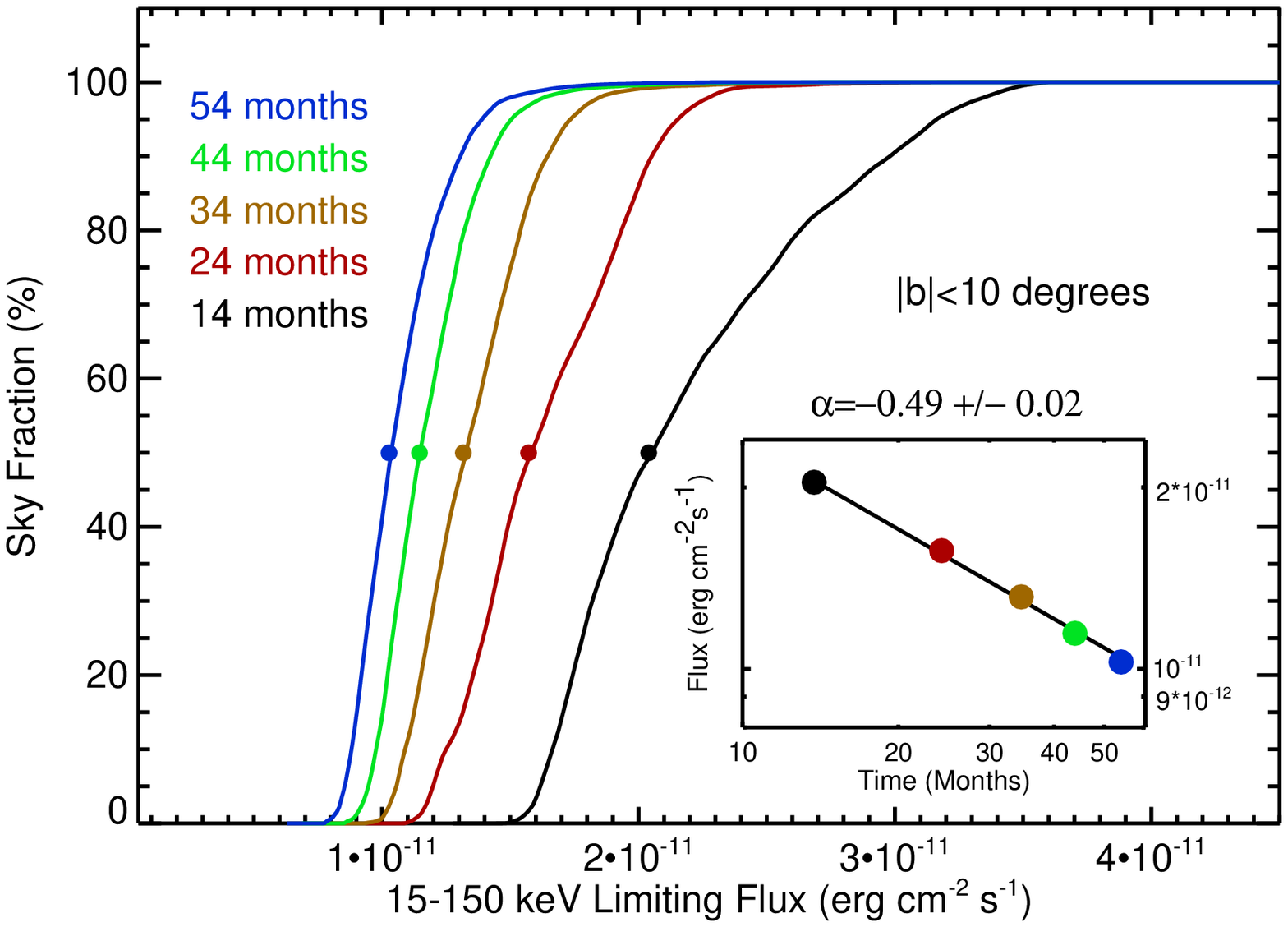,width=9cm,angle=0}}
\centerline{\psfig{figure=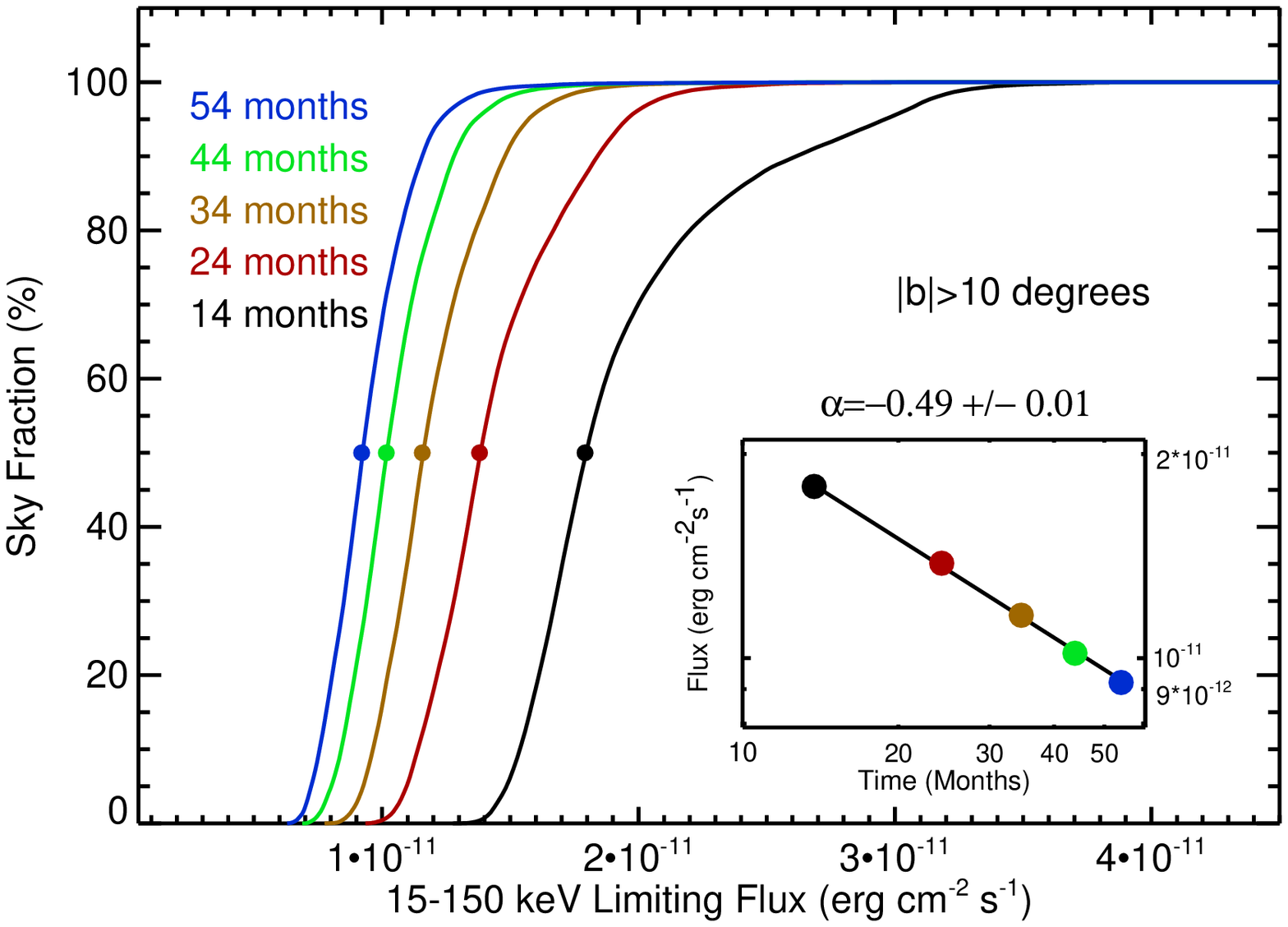,width=9cm,angle=0}}
\caption{Fraction of the sky (top: $|{\rm b}|< 10^{\circ}$, bottom: $|{\rm b}|> 10^{\circ}$) covered by 
the BAT survey 
as a function of the 15--150 keV detection limiting flux for a detection threshold of 4.8 standard
deviations. 
Different colours refer to different survey epochs. The insets show the 
 limiting flux achieved for 50\% of the sky as a function of time; the best fit 
is a power law consistent with t$^{-0.5}$.
\label{skycov}}

\end{figure}

Figure~\ref{skycov} shows the sky coverage, defined as the fraction of the sky 
covered by the 
survey as a function of the 15--150 keV detection limiting flux, at different survey epochs 
starting from the beginning of the mission. 
 The limiting flux of a given 
sky direction is calculated  by multiplying the local image noise by a 
detection threshold of $4.8$ standard deviations. 
We derived the sky fraction for two sky regions (top panel: $|{\rm b}|< 10^{\circ}$, 
bottom panel: $|{\rm b}|> 10^{\circ}$). 
The 54-month BAT survey covers 50\% of the sky to a
flux limit of  $1.0\times10^{-11}$ \ferg\ and $9.2\times10^{-12}$ 
\ferg\ for $|{\rm b}|< 10^{\circ}$  and $|{\rm b}|> 10^{\circ}$, respectively.
The insets in Fig.~\ref{skycov} show the limiting flux achieved for 50\% 
of the sky as a function of the  cumulative observing time of the screened BAT survey data; 
the data are modelled well with a power law 
(N$\times$t$^{\alpha}$, where N=$(7.5\pm 0.3)\times 10^{-11}$ \ferg and  $\alpha=-0.49 \pm 0.02$ 
 for $|{\rm b}|< 10^{\circ}$  and N=$(6.4\pm 0.2)\times 10^{-11}$ \ferg and  $\alpha=-0.49 \pm 0.01$ for 
$|{\rm b}|> 10^{\circ}$), both being consistent with the 
t$^{-0.5}$ behaviour expected if the statistical errors dominate over the 
systematic ones.

\begin{figure*}
\centerline{\psfig{figure=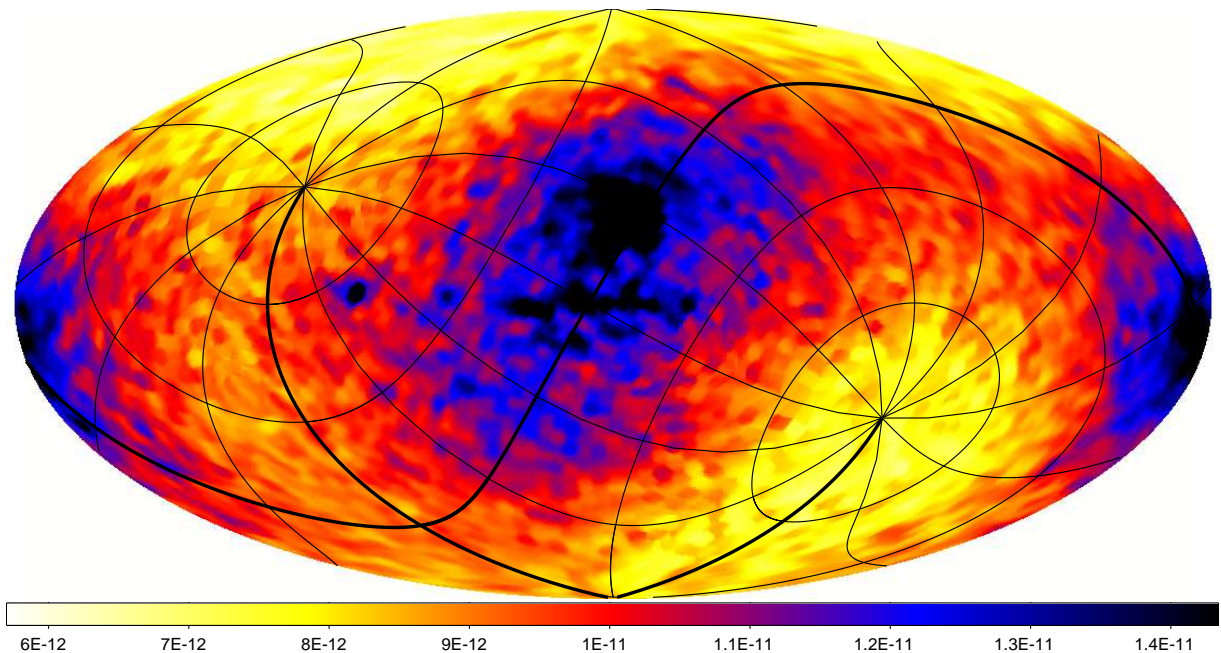,width=16cm}
            }
\caption{Map of the limiting flux of the 54-month BAT survey in the 15--150 keV 
band, projected in Galactic Aitoff coordinates, with the ecliptic coordinates grid superimposed. The
colour bar shows the scale in \ferg.
\label{limitfl}}
\end{figure*}

The minimum detection limiting flux is not fully uniform on the sky. Figure~\ref{limitfl} shows the 
limiting flux map in Galactic Aitoff projection, with the ecliptic coordinates grid 
superimposed. The Galactic centre 
and the ecliptic plane are characterized by a poorer sensitivity because of high contamination from 
intense Galactic sources and to the observing constraints on the {\it Swift} spacecraft. The highest flux 
sensitivity is achieved close to the ecliptic poles, where a detection flux limit of 
$\sim6.2 \times 10^{-12}$ \ferg\ is reached; the lowest flux sensitivity is in 
the region of the Galactic centre with a detection flux limit of $\sim3 \times 10^{-11}$ \ferg.

We produced all-sky maps in three energy bands (15--30 keV, 15--70 keV and 15--150 keV) using 
the HEALPIX-based all-sky spherical grid  projection \citep{HEALPix} with a pixel size 
of $\approx$2.5 arcmin radius. 

\begin{figure}
\centerline{\psfig{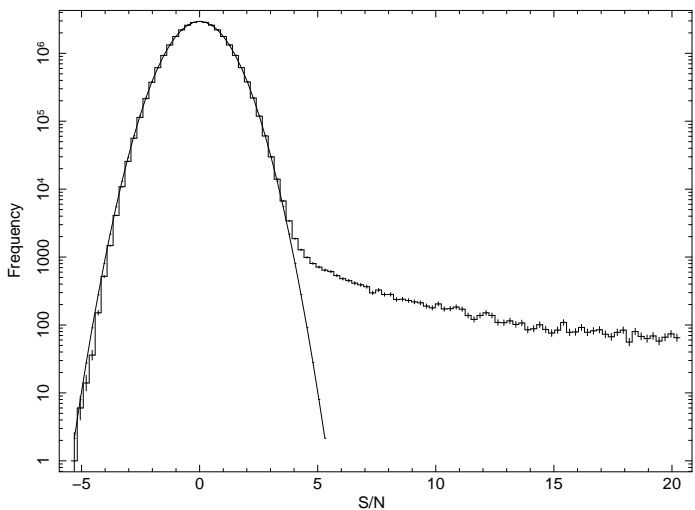}
            }
\caption{Distribution of pixel significance in the BAT all-sky map. The continuous curve is the
result of a Gaussian fit obtained by excluding the distribution tail. The best-fit model 
parameters are consistent with a mean and standard deviation of 0.0 and 1.0, respectively. 
\label{sigma}}

\end{figure}

For each of these energy ranges, we derived a signal-to-noise ratio
(S/N) map as the ratio of the mosaic intensity to the associated statistical error. 
Figure~\ref{sigma} shows the distribution of the significance in the 15--150 keV energy range. 
This distribution is well described by a Gaussian curve with zero mean and unitary variance, 
except for the positive tail caused by hard X-ray emitters. The same result was also obtained 
for the significance maps in the other two energy ranges.

\section{Detection strategy\label{detstrategy}}
The source detection was performed by searching
for local excesses in the S/N maps, then refining their position and peak significance
using a local bidimensional fit.
Detections with peak significance greater than 
4.8 standard deviations were included in our list of detected sources. 


Adopting this threshold, we expected $\sim15$ spurious detections on each all-sky map: this 
number was evaluated by applying the detection algorithm to several all-sky maps obtained 
from simulated empty field observations. Therefore, the total number 
of spurious detections was between 15 and 45 ($\sim1\%$ to $\sim3\%$ of the total number
of our detections, see below), the best case occurring if each
noise fluctuation above the threshold appeared simultaneously in all the three 
bands, the worst case occurring if each fluctuation appeared only in one energy 
band. 

The final catalogue is built by cross-correlating and merging the detection 
catalogues obtained in the three energy bands: source candidates detected in the 
sky maps of different energy bands were merged and reported in the final 
catalogue as a single source candidate if their positions were consistent
within the relevant error box (95\% containment radius, \citealp{segreto10}).

We obtained a list of 1256 source candidates detected in at least
one of the three energy bands: 
806 sources were detected in all the three energy bands, 230 in two energy bands, and
220 in only one of the three energy bands (74, 59, and 87 in the 15--150 keV, 15--30 
keV, and 15--70 keV map, respectively). We assume the most accurate source coordinates
to be those obtained in the sky map with the highest detection
significance.

\section{Association strategy\label{id}}

To find the most likely counterpart to the detected BAT hard X-ray
excesses, we applied two different strategies: an analysis of 
archival soft X-ray observations (strategy A) and a cross-correlation with a 
list of possible counterparts (strategy B).

\subsection{Strategy A} We analysed all the available soft X-ray archival observations 
whose field of view covers the position of the BAT source candidates. 
We first considered the huge set of \sw-XRT observations, many of which were 
performed for this purpose. A total of 751 sky positions of the BAT source 
candidates were 
covered by XRT observations. We applied a blind detection algorithm  to the XRT
images using {\sc ximage} v4.0. We assumed that an XRT source was the counterpart of
a BAT detection if its position was within a 6 arcmin radius error circle
(99.7\% confidence level for a source detection at 4.8 standard deviations,
\citealp{segreto10}) and its rate was higher than $8\times 10^{-3}$ count~s$^{-1}$ 
in the 0.2-10 keV energy range or higher than $8\times 10^{-4}$ count~s$^{-1}$ in 
the 3-10 keV energy range (criterion 1). These two thresholds were derived by assuming 
that a source is detected at about the survey limiting flux ($\sim10^{-11}$ \ferg, see 
Fig.~\ref{skycov}) and  extrapolating the XRT count rate to a power-law spectral 
energy distribution of photon index $1<\Gamma<3$ and an absorbing column 
$\rm 10^{20}$ cm$^{-2} <Nh<10^{24}$ cm${-2}$ and allow us to associate either faint or 
very absorbed sources with the BAT detection. We found that 595 BAT excesses 
could be associated with a 
single XRT source, while 60 BAT excesses could be associated with more than one XRT 
source (42 with a double association and 18 with a triple association).
In the latter cases, we associate to the BAT excess the XRT source with either a 0.2-10 keV 
or 3-10 keV count rate at least a factor of 5 brighter than the other candidates in 
the field  (criterion 2). This criterion leaves only 8 BAT excesses with a double 
XRT association, which are reported in the catalogue. The number of XRT counterpart 
candidates rejected after applying of this criterion is $42+18\times2-8=70$.
For 96 of the BAT source candidates covered by an XRT observation, we were 
unable to detect any soft X-ray counterpart. 

To evaluate the
number of expected spurious associations, we collected a large sample (365) of XRT
observations of GRB fields, using only late follow-ups (where the GRB afterglow
had faded) with a similar exposure time
distribution as the XRT pointings of the BAT sources. We searched for
sources within a 6.0 arcmin error circle centred on the nominal pointing
position in each of these fields (excluding any GRB residual afterglow) and
satisfying criterion 1. We detected 33 sources that, normalized to the number 
of XRT follow-ups ($33\times751/365\simeq68$), is consistent with the number 
(70) of XRT sources that survived criterion 1 but were rejected by criterion 2.
Therefore, the number of expected spurious associations can be assumed to be negligible.

For the BAT positions not covered by XRT observations, we searched for pointed archival
observations with other X-ray instruments, in the following order: Beppo-SAX,
ASCA, Newton-XMM, Chandra, ROSAT. We did not use ROSAT observations performed
during the ROSAT All Sky Survey campaign: the list of sources extracted from this campaign
\citep{voges99} was used in strategy B (see Sect.~\ref{B}).
A threshold criterion analogous to that
applied to the \sw-XRT observations was used to select the most reliable
association. The rate thresholds for criterion 1 for each 
instrument were derived by converting the \sw-XRT count rate threshold to the relevant 
equivalent count rate assuming a power law with a photon index $\Gamma=2$ and an absorbing 
column of $10^{21} \rm cm^{-2}$.
For ROSAT, only a 0.2--2.4 keV rate threshold was applied. 
A total of 288 of the BAT source candidates positions not observed with \sw-XRT were 
covered by observations of these other X-ray telescopes. 
We identified 275 unambiguous associations and 5 double possible associations.
To resolve the ambiguity in these cases, we applied  criterion 2 as for XRT,
finding  no BAT excesses for more than one possible source association. 
Since we applied the same threshold criteria as for \sw-XRT, we can confidently assume a negligible 
number of spurious associations.

Finally, the identification of the soft X-ray counterpart was performed by searching in the 
SIMBAD\footnote{http://simbad.u-strasbg.fr/simbad/} and NED\footnote{http://nedwww.ipac.caltech.edu/}
databases within the soft X-ray error box. In the few cases where the soft X-ray counterpart is 
an unknown source,
we report it in our catalogue as a new source with a name composed by the PBCX 
acronym (Palermo BAT Catalogue X-ray source) followed by its soft X-ray coordinates with the precision of 
1.5 arcsec in RA and 1 arcsec in Dec.

With strategy A we were able to associate 920 BAT excesses to a single softer counterpart and 8 BAT excesses to a double softer counterpart.
328 BAT excesses still lacking an association.

\begin{figure}
\centerline{\psfig{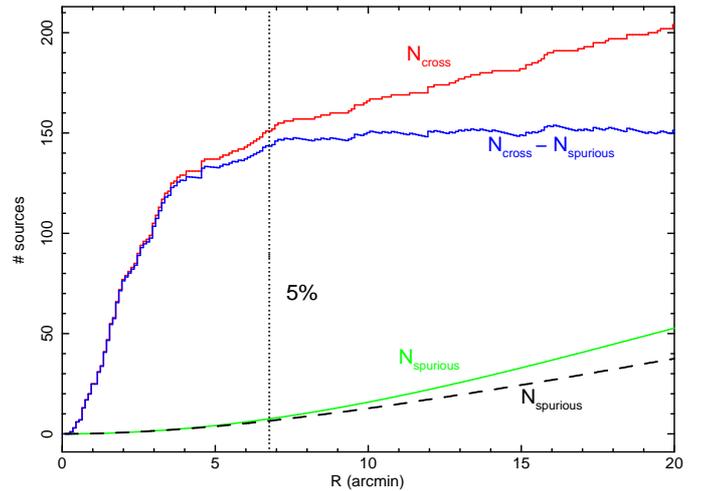}
            }
\caption{Cumulative distribution of the number of BAT excesses not associated with strategy A 
having at least one of the strategy B list sources 
(see Sect~\ref{id}) within a given distance (red stepped line). 
The green continuous line represents the  number of spurious associations evaluated using
Eq.~\ref{sp}, while the black
dashed line is the mean number of spurious associations evaluated by using a control sample of sky positions generated 
by scrambling the coordinates of the BAT excesses. 
The blue stepped line represents the true associations obtained as the difference between the red stepped line and the green continuous line. 
The vertical dotted line marks the radius (6.8 arcmin) that produces 5\% of spurious associations.
\label{simbad}}

\end{figure}

\subsection{Strategy B \label{B}} 
To find an association for the 328 BAT excesses still lacking an association,
we adopted the following strategy. 
We compiled a list of possible counterparts (hereafter strategy B source list: 
SBSL) merging the following catalogues: 
\begin{itemize}
\item high and low mass X-ray binaries, cataclysmic variables, supernova remnants 
 and pulsars, Seyfert galaxies, unclassified AGNs, cluster of galaxies, interacting galaxies, 
 LINERs, and  $\gamma$-ray sources, 
 whose lists were extracted from the SIMBAD database on January 2010;
\item the {\it Roma}-BZCAT \citep{massaro09};
\item the ROSAT All Sky Survey (RASS) Bright source catalogue \citep{voges99}.
\end{itemize} 
\noindent The resulting catalogue contains ${\rm N_{SBSL}} = 60$ 829 sources.

The number ${\rm N_{cross}(R)}$ of BAT excesses for which at least one SBSL source 
was within a specified distance R is represented by the red stepped line in 
Fig.~\ref{simbad}. 

Assuming that $\rm N_{true}$ of the  $\rm N_{BAT}= 328$ BAT excesses 
have a counterpart in a generic catalogue  of                            
${\rm N_{cat}}$  sources evenly distributed across the sky with a density
$\lambda={\rm N_{cat}}/4\pi$, the  number of 
expected spurious associations ${\rm N_{spurious}}$ generated by the $\rm N_{BAT} - \rm N_{true}$
sources without a counterpart in the catalogue is expressed by

\begin{equation}
\label{sp}
{\rm N_{spurious}(R)} = \left({\rm N_{BAT}} - {\rm N_{true}}\right) \times \left(1-e^{-{\pi\lambda R^2}}\right).
\end{equation}

\noindent To a first approximation we assumed that SBSL is uniformly distributed across the sky,
so we apply the above expression with $\rm N_{cat} = \rm N_{SBSL}$. 
Since $\rm N_{true}$  is not known in advance, we used the following procedure: we 
increased  $\rm N_{true}$ with a unitary step and evaluated the 
difference between  ${\rm N_{cross}(R)}$ and $\rm N_{spurious}(R)$.
after increasing the correlation radius, this curve flattens because no further true 
associations are obtained. This happens for $\rm N_{true} \sim 195$. The blue 
stepped line in Fig.~\ref{simbad} shows $\rm N_{cross}(R) - N_{spurious}(R)$
and the green continuous line represents $\rm N_{spurious}(R)$.

As a further check, we defined a control sample, by generating 1000 lists of 
$328 - \rm N_{true} = 133$ sky positions: to preserve the Galactic coordinate 
distribution we scrambled the arrays of Galactic latitude and longitude of the BAT 
coordinate excesses and then extracted 133 couples of coordinates from these 
scrambled arrays. The mean number of spurious associations was then evaluated as a 
function of the association radius (Fig.~\ref{simbad}, dashed line). This curve is 
in perfect agreement with the analytical one (green continuous line) out to $\sim$ 
8 arcmin and increases in size more slowly at larger distance. We verified that this difference 
is due to the inhomogeneity in SBSL and in particular to the clustering of sources 
in regions covered by deep optical surveys.  

\begin{figure}
\centerline{\psfig{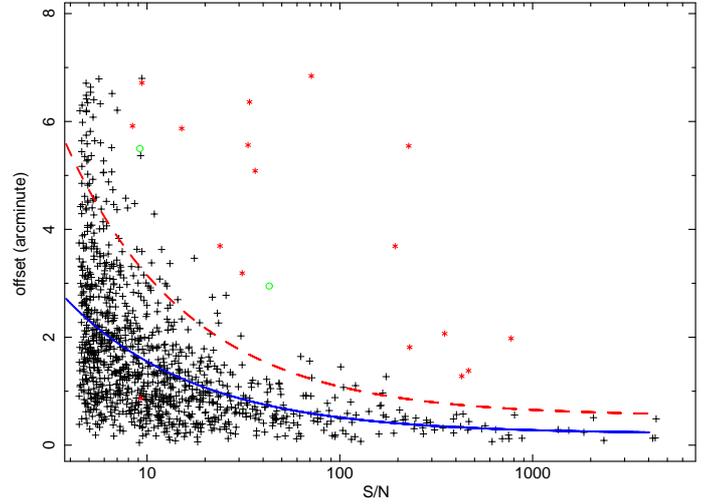}
            }
\caption{Offset between the BAT position and the counterpart position as 
a function 
of the detection significance. 
A few values are far from the overall
distribution either because they are in crowded fields and their reconstructed sky position
is contaminated by the PSF of the nearest sources (red stars) or because
they are extended sources (green circles).
The solid blue line represents the fit to the data (excluding the outliers) with      
a power law. The dashed red line represents the  95\% containment radius.
\label{offset}}

\end{figure}

The ratio of $\rm N_{spurious}(R)$ to ${\rm N_{cross}(R)}$ is an estimate 
of the fraction of spurious association as a function of the association radius. We 
decided to accept a maximum of 5\% of spurious associations that correspond to an
association radius of 6.8 arcmin. 
With this strategy, we associate 151 BAT sources with counterparts 
(131 with a single counterpart, 18 with a double 
counterpart, and 2 with a triple counterpart).
The expected  number of spurious associations is 6.8 $\pm$ 2.5.
\\
\\
As a result of these two association procedures, we found that 1079 BAT sources have 
at least an associated counterpart 
(1051 with a single counterpart, 26 with a double counterpart, and 2 with  a triple counterpart)
and that 177 sources still lack a counterpart. The probability of spurious association is negligible 
for sources associated with strategy A 
and 5\% for those associated with strategy B. \\       
\\
In Fig.~\ref{offset}, we plot the offsets of each BAT source excess with respect
to its associated counterpart versus (vs.) the detection
significance (S/N). 
The offset of a few sources is far from the overall 
distribution: points indicated by a star 
(sources number 286, 328, 796, 856, 860, 869, 870, 902, 928, 949, 
950, 954, 955, 956, 963, 977, and 979  in Table 2) are 
in crowded fields, and the reconstructed sky position is 
contaminated by the PSF of the nearest sources; those marked with 
a circle are extended sources (Coma cluster and Abell 2256). 
The distribution (excluding the outliers) can
be modelled with a power law plus a constant giving the following best fit 
equation:
\begin{equation}
{\rm Offset(')} = (7.2\pm 1.2)\times {\rm [S/N]}^{-0.76\pm0.10} +(0.21\pm0.03),
\end{equation}
where the constant  represents the systematic offset. At the detection threshold of
4.8 standard deviations, the average offset is  2.4 arcmin.
The dashed red line in Fig.~\ref{offset} represents the 95\% radius 
containment radius evaluated as described in \citet{segreto10},
\begin{equation}
\rm R_{95}(') = 12.5 \times [S/N]^{-0.68} + 0.54 
\end{equation}
where S/N is the detection significance.


\section{The 54-month catalogue\label{cat}}

The complete catalogue of the sources detected in the first 54
months of BAT survey data is reported in Table~\ref{srctab}.
The table contains the following information:
\begin{itemize}
\item Second Palermo BAT catalogue (2PBC) name of the source (Col. 2), built from the BAT coordinates 
with the precision of 1.5 arcmin on RA and 1 arcmin on Dec.
\item Counterpart association (Col. 3) and source type (Col. 4) 
coded according to the nomenclature used in SIMBAD. 
For the blazars included in  the Roma-BZCAT \citep{massaro09}, we report the 
nomenclature used in that catalogue: BZB for BL Lac objects, BZQ for flat-spectrum 
radio quasars, and BZU for blazars of uncertain type.
\item The RA and Dec of the BAT source in decimal degrees (Cols. 5, 6) measured in the energy 
band with the highest detection significance.
\item The 95\% error radius (Col. 7) and offset with respect to the counterpart position 
(Col. 8).
\item Source significance (Col. 9) obtained in the energy band with the 
highest significance (a flag in Col. 19 indicates the energy range with the 
maximum significance).

\item Flux and errors (Cols. 10 and 11) in the 15--150 keV band averaged over the entire
survey period.
To produce spectra for the detected sources, we created all-sky maps in eight energy bands 
(15--20 keV, 20--24 keV, 24--35 keV, 35--45 keV, 45--60 keV, 60--75 keV, 75--100 keV, and 100--150 keV) 
from which we extracted the rates and their errors from the pixel corresponding to the most 
likely position of each BAT source  (Sect.~\ref{detstrategy}). 
These spectra were analysed using the BAT spectral redistribution
matrix\footnote{http://heasarc.gsfc.nasa.gov/docs/heasarc/caldb/data/swift/bat/index.html} and the fluxes in the
15--150 keV were evaluated by fitting the spectra with a simple power law.
 
\item Hardness ratio (HR, Col. 12)  and error (Col. 13) obtained as the ratio of the counts in the 35--150 keV band 
to those in the 15--150 keV band. 

\item Redshift of the extragalactic sources (Col. 14) from the SIMBAD
database (or NED, for the few cases that were not reported in SIMBAD).
\item Rest-frame luminosity (in units of log[erg s$^{-1}$])
in the 15--150 keV band (Col. 15) calculated, when the redshift is available, using the expression
\begin{equation}
L_{15-150 keV}=4 \pi D_L^2 \frac{F_{15-150}}{(1+z)^{2-\Gamma}},
\end{equation}
where $F_{15-150}$ is the observed flux in the 15--150 keV band, $\Gamma$ is the photon index
obtained from the spectral fit, $D_L$ is the luminosity distance of the source, and $z$ 
is its redshift.
For sources with redshift $<0.01$, we used the distance
reported in the Nearby Galaxies Catalogue (NBG, \citealp{tully88}) or NED, for 
the few cases that were not reported in the NBG catalogue.

\item Variability index (Col. 16). In this second catalogue, we added a characterization 
of the time behaviour of the BAT-detected sources: 
the light curve of each source was binned at 7 days and the variability was investigated using a
simple $\chi^2$ test. The rate in the jth 7-day time bin ($R_j$) is evaluated by weighting the rates of the light curve at maximum resolution by
the inverse square of the corresponding statistical error 
\begin{equation}
R_j=\frac{\sum{r_i/er_i^2}}{\sum{1/er_i^2}},
\end{equation}
where  $r_i$  are the rates observed in the light curve at 
maximum resolution, and $er_i$ are the corresponding statistical errors. The error in 
$R_j$ is $ER_J=\left(\sqrt{\sum{1/er_i^2}}\right)^{-1}$.
The variability index is defined as 
\begin{equation}
V=\sum{w_j(R_j-<R>)},
\end{equation}
where $w_j=[ER_J^2+(f\times R_j)^2]^{-1}$ and $<R>={\sum{w_j R_j}}/{\sum{w_j}}$. A systematic 
error of $f\times R_j$ 
with $f=5\%$  was added in quadrature to the statistical error of each bin, to obtain a 
variability index V$\sim 1$ for Crab, Vela Pulsar, and PSR 0540-69.

\item Flag column (Col. 17) with information on: energy band with the highest
significance (A), flag for already known hard
X-ray sources (B), position with respect to the Galactic plane (C), and strategy 
used for the identification (D, see Sect.~\ref{id})

\item Flag column (Col. 18) with information on the cross correlation between the BAT sources and the ROSAT, 
INTEGRAL, and {\it Fermi} catalogues.
A BAT source is associated with a ROSAT source if the BAT counterpart lies within the 3$\sigma$  error box of a source reported  
in the RASS bright and faint source catalogues \citep{voges99, voges00}. 
The cross-correlations of the BAT catalogue with the ISGRI sources and the Fermi sources were performed 
using the INTEGRAL General Reference Catalogue 
V.31\footnote{http://www.isdc.unige.ch/integral/data/catalog} and the Fermi Large Area Telescope First Source Catalogue
\footnote{http://fermi.gsfc.nasa.gov/ssc/data/access/lat/1yr\_catalog/}\citep{abdo10}, respectively, requiring that the
sources had the same associated counterpart.
\end{itemize}

\section{Statistical properties of the catalogue\label{catprop}}

\begin{figure*}
\centerline{\psfig{figure=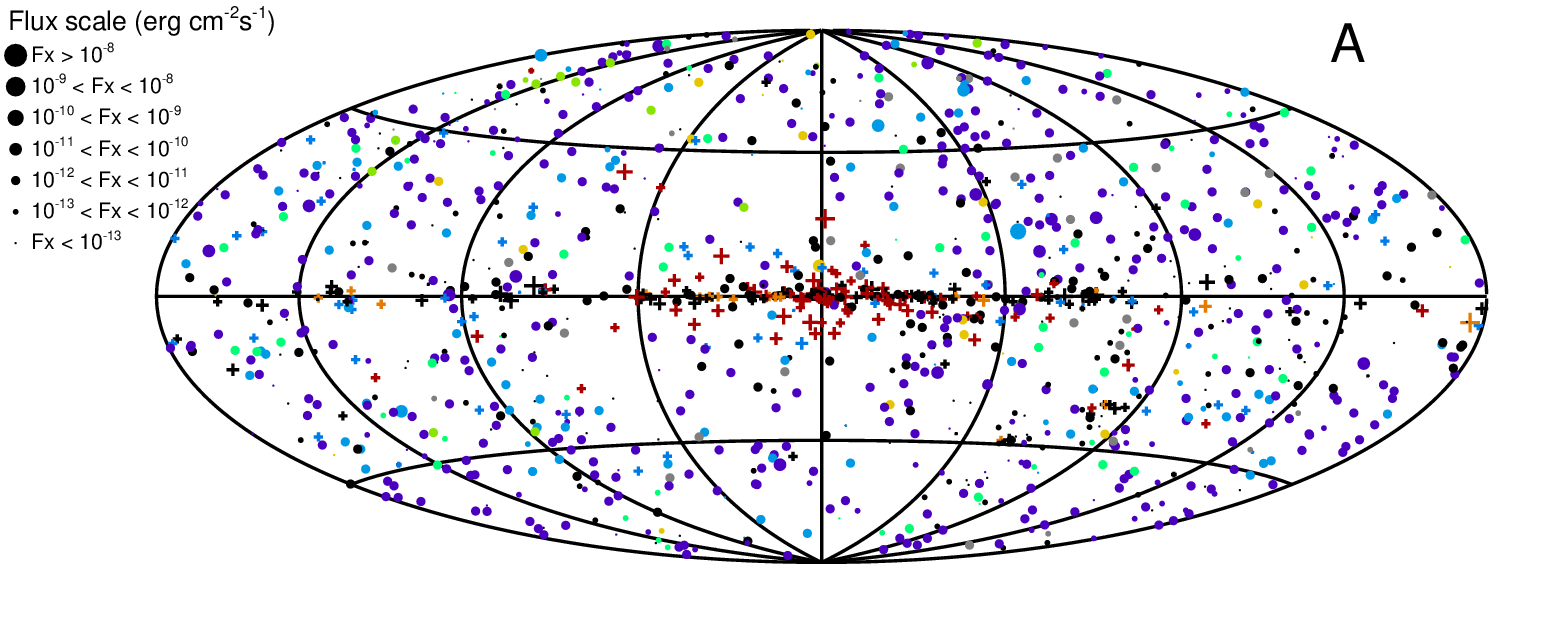,width=16cm}
            }
	    \centerline{\psfig{figure=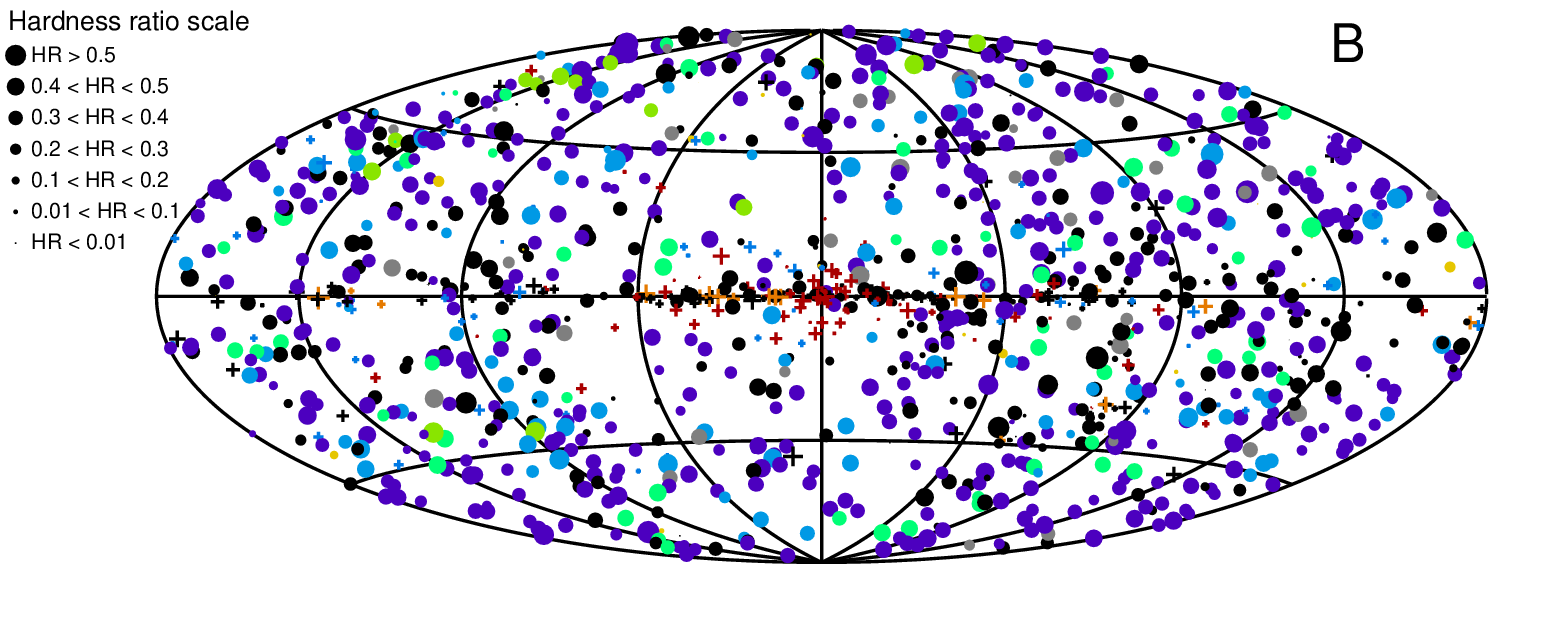,width=16cm}
            }
	    \centerline{\psfig{figure=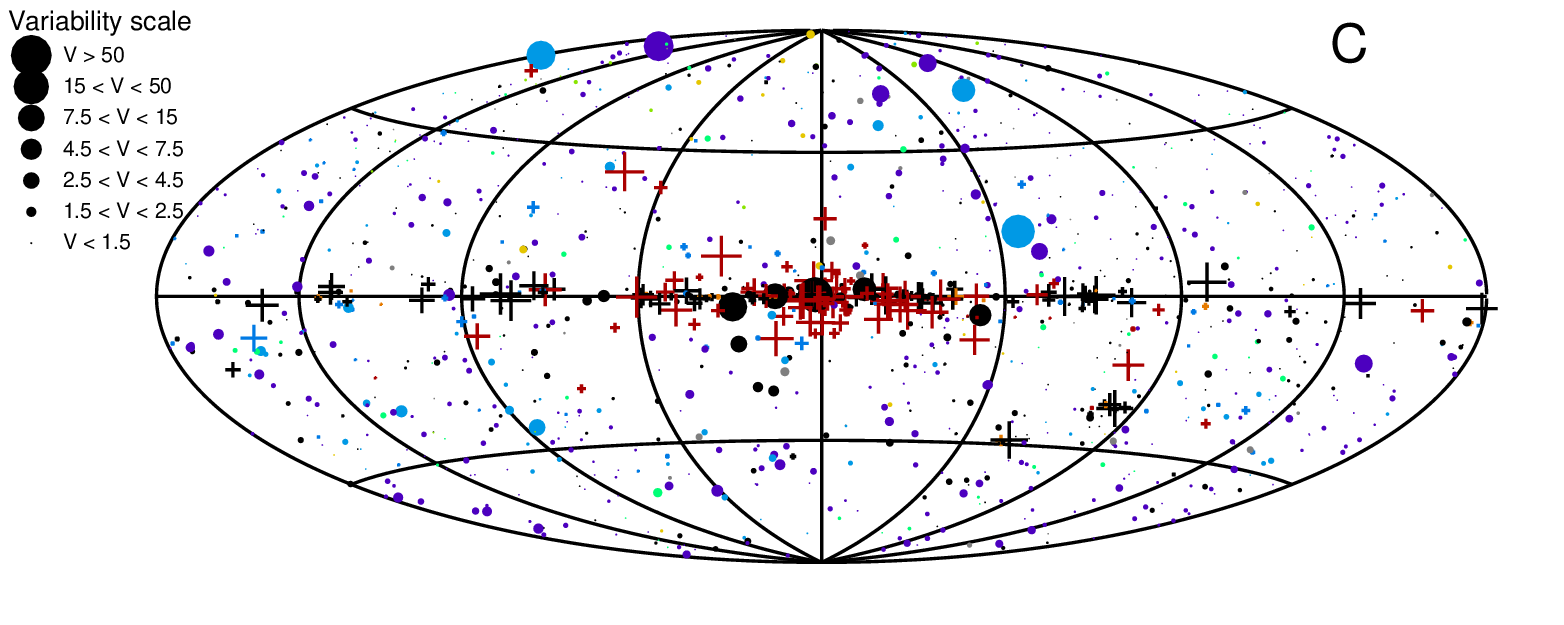,width=16cm}
            }
	    \centerline{\psfig{figure=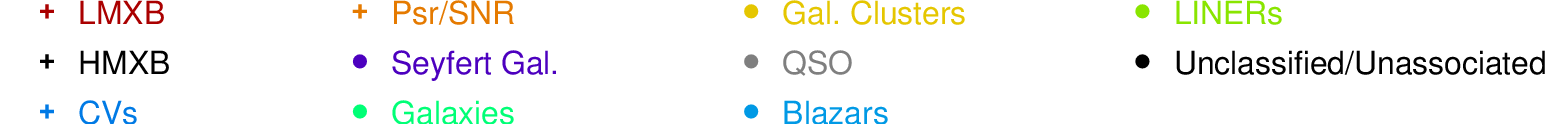,width=16cm}
            }
\caption{Map of the sources (in Galactic coordinates) detected in the BAT survey data. The object
class is colour-coded according to the legend.   The size of the symbol is proportional to {\bf (A)} the
15--150 keV source flux, {\bf (B)}  the hardness ratio  obtained as the ratio of 
the counts in the 35--150 keV band to those in the 15--150 keV band, {\bf (C)}  the variability index 
(as defined in Sect.~\ref{cat}).
\label{aitoff}}
\end{figure*}

Table~\ref{types} compares the numbers of counterparts associated with the 
sources detected in the 54-month all-sky mosaic 
among the different object classes, with similar results for the 39-month
catalogue. Percentages are evaluated for both catalogues relative to the 
total number of BAT-detected sources.
The sample consists of $\sim19\%$ Galactic sources, $\sim56\%$ extragalactic sources, and
$\sim10\%$ sources with a counterpart at softer energies whose nature has not yet been determined.
We also found that $\sim15\%$ of sources have no association at other wavelengths.
The distribution of the associated sources among the different classes is almost identical to  
that of the 39-month catalogue.
There is a significant difference for the fraction of unassociated sources, which is a 
factor $\sim1.6$ lower than in the 39-month catalogue. This is because a \emph{Swift}-XRT 
follow-up campaign was 
requested for the unassociated sources of the 39-month catalogue and  the ROSAT 
catalogue was used in the association strategy of the 54-month catalogue. In contrast, we 
have a much higher
fraction of unclassified sources ($\sim10\%$), most of which are ROSAT sources 
of unknown nature. Figure~\ref{aitoff} shows the map of the detected sources, colour-coded according to 
the object class and size-coded according to the 15--150 keV source flux (A), the hardness ratio 
(B), and the variability index (C), respectively.

\begin{table}
\begin{tabular}{l r r}
\hline\hline
Class                 & \# in 54m (\%)& \#  in 39m (\%)\\ \hline
LXB                   & 85 (6.6)       &  76  (7.9)  \\
HXB                   & 83 (6.5)	 &  64 (6.6)\\
Pulsars               & 11 (0.9)	 &  10 (1.0)\\
SNR                   & 7   (0.5)	 &  5 (0.5)\\
Cataclysmic variables & 56 (4.4)	 &  46 (4.8)\\
Stars                 & 7   (0.5)	 &  5 (0.5)\\
Star clusters         & 1   (0.1)	  &   0 (0.0) \\ \hline
Galactic (total)      & 250 (19.5)	 &  207 (21.5)\\  \hline  
Seyfert 1 galaxies    & 307  (23.9)	 & 235 (24.4)\\ 
Seyfert 2 galaxies    & 165  (12.8)	 & 131 (13.6)\\ 
LINERs                & 15    (1.2)	 & 7  (0.7)\\
QSO                   & 25   (1.9)	 & 14  (1.5)\\
Blazars               & 97   (7.5)	 & 71  (7.4)\\
Interacting galaxies  & 2    (0.16)       &  0   (0.0) \\
Galaxy clusters       & 23   (1.8)	 & 18  (1.9)\\
Normal galaxies       & 67   (5.2)	 & 27  (2.8)\\
Unclassified AGN      & 34   (2.6)	& 16 (1.7)\\ \hline
Extragalactic (total) & 735  (57.1)	 & 519 (54.0)\\ \hline 
Unclassified sources  & 124  (9.6)	 & 28 (2.9)\\ 
Unassociated sources  & 177  (13.8)	  & 208 (21.6)\\ \hline  
\end{tabular}
\caption{Classification of the counterparts associated with the sources detected in the 
54-month BAT  survey. 
{\it Unclassified sources} includes all sources that have a catalogued counterpart but have not
yet been classified. 
\label{types}}
\end{table}

\begin{figure*}
\centerline{\psfig{figure=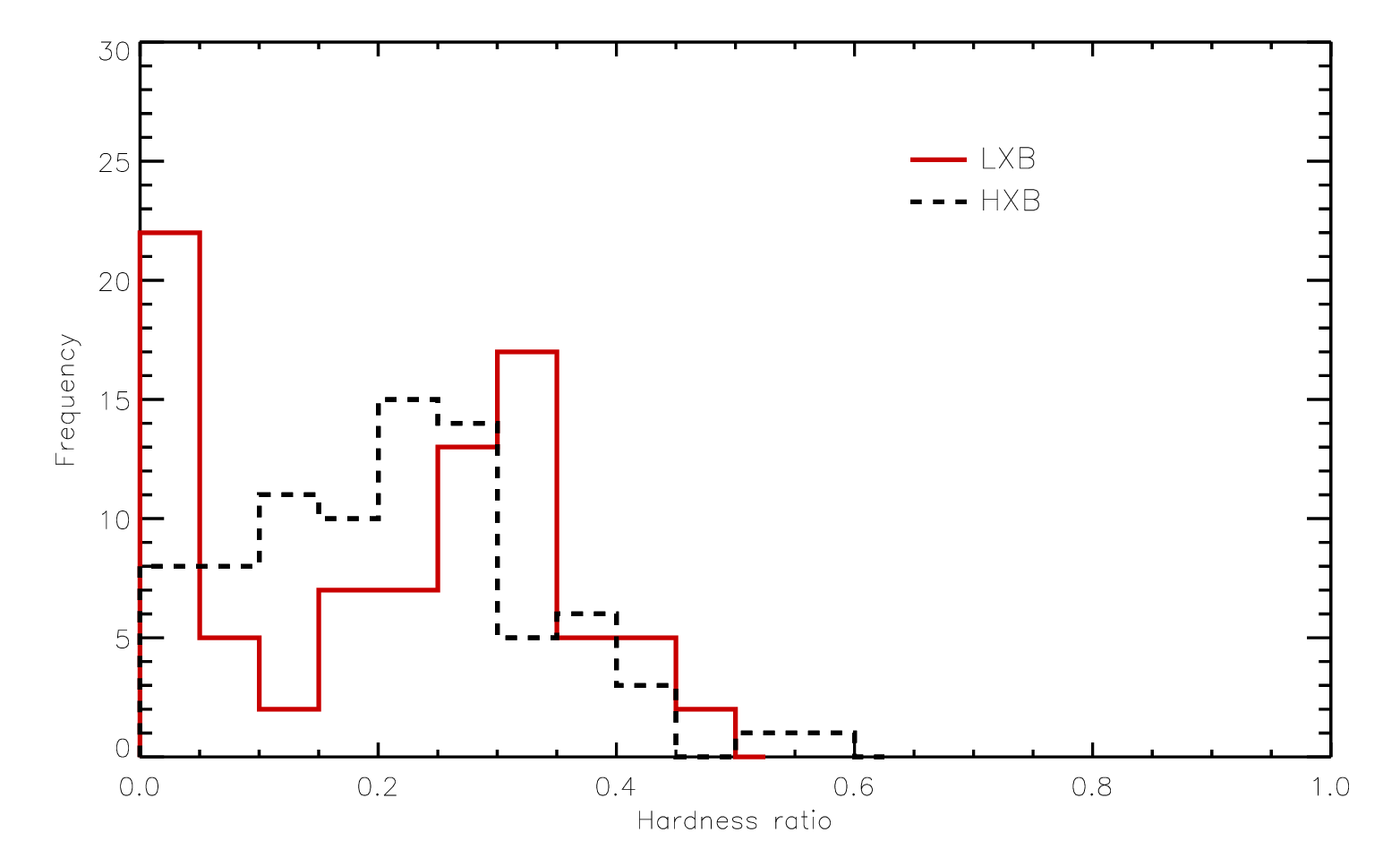,width=8.5cm,angle=0} \psfig{figure=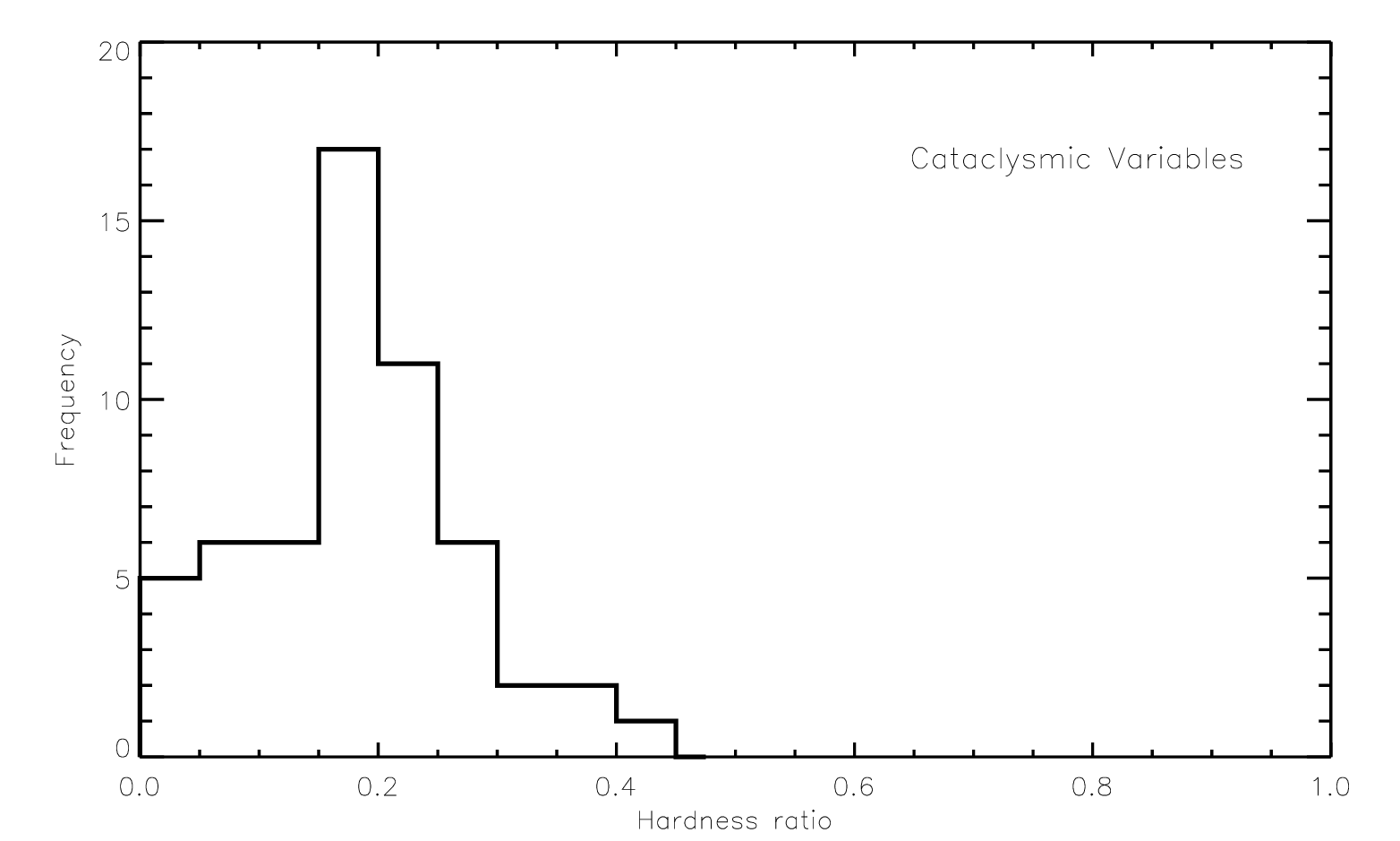,width=8.5cm,angle=0} }
\centerline{\psfig{figure=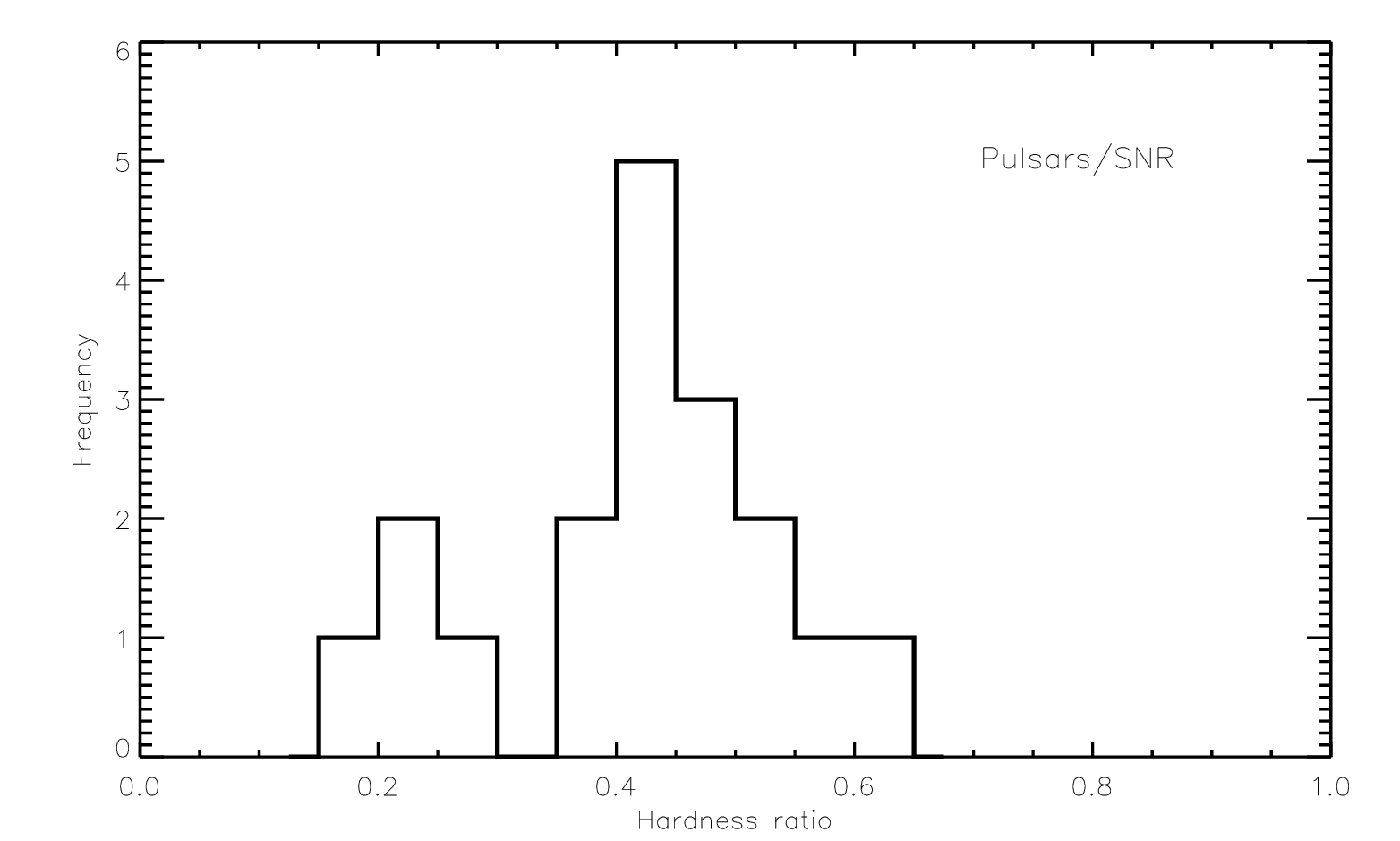,width=8.5cm,angle=0} \psfig{figure=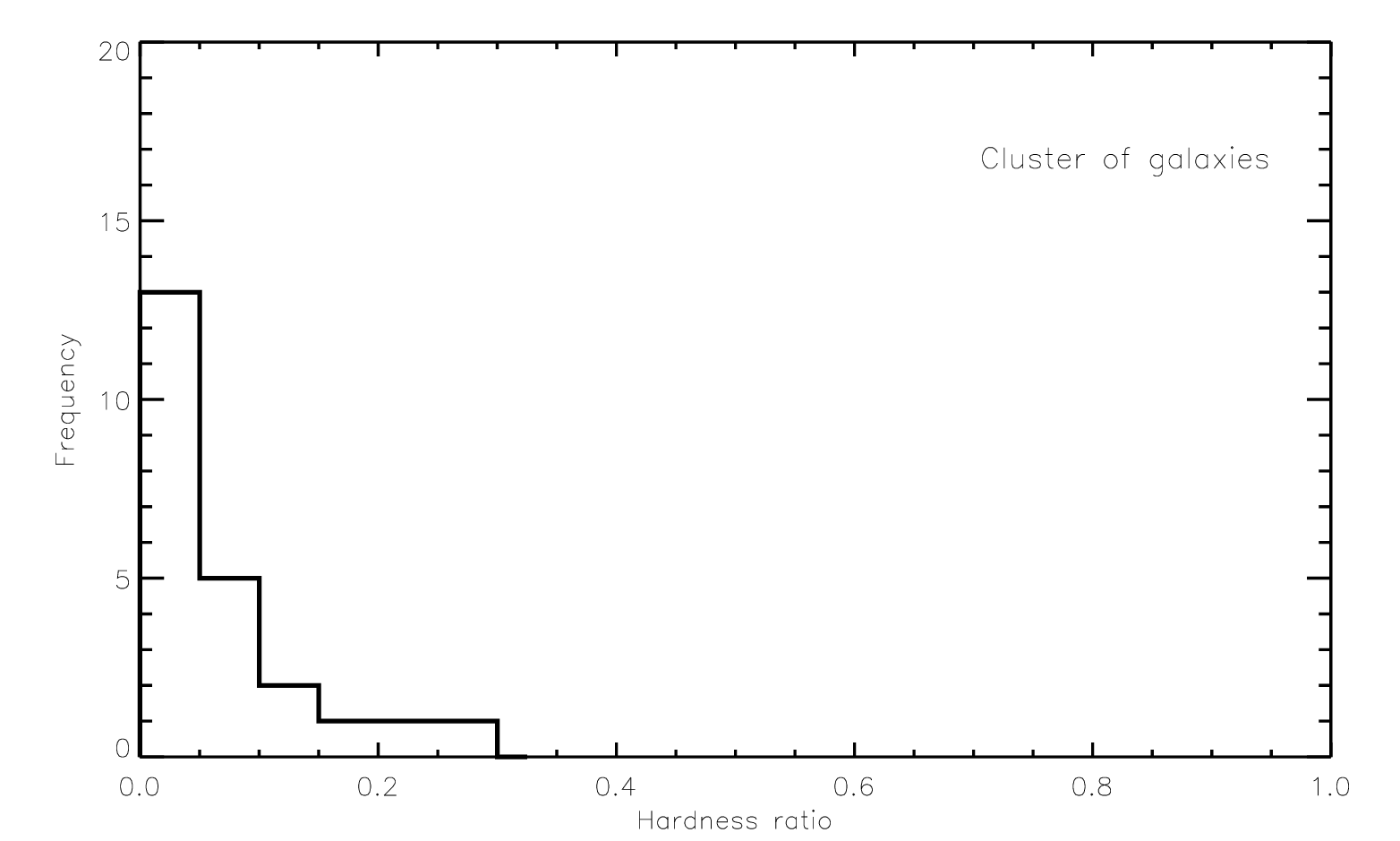,width=8.5cm,angle=0} }
\centerline{\psfig{figure=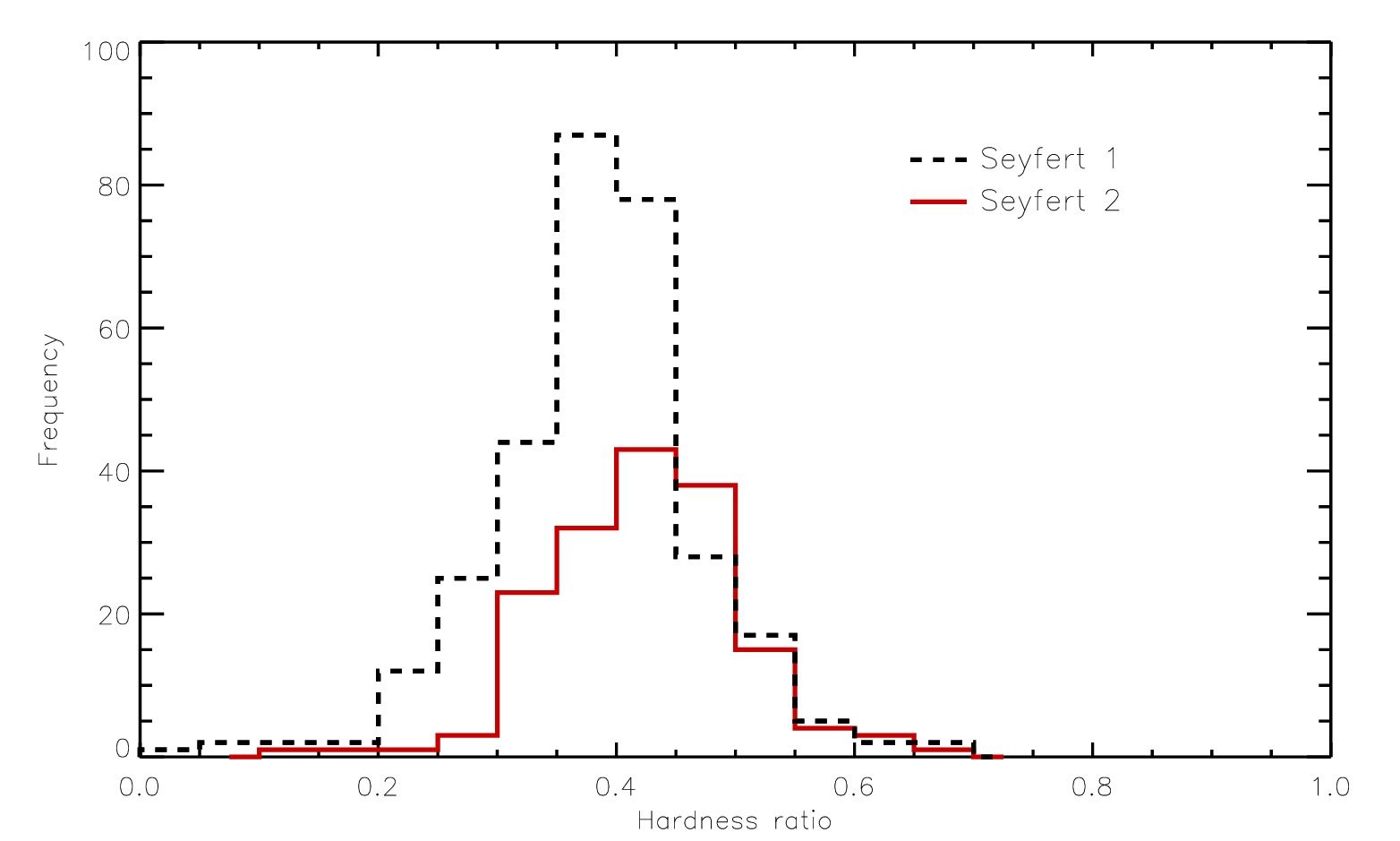,width=8.5cm,angle=0} \psfig{figure=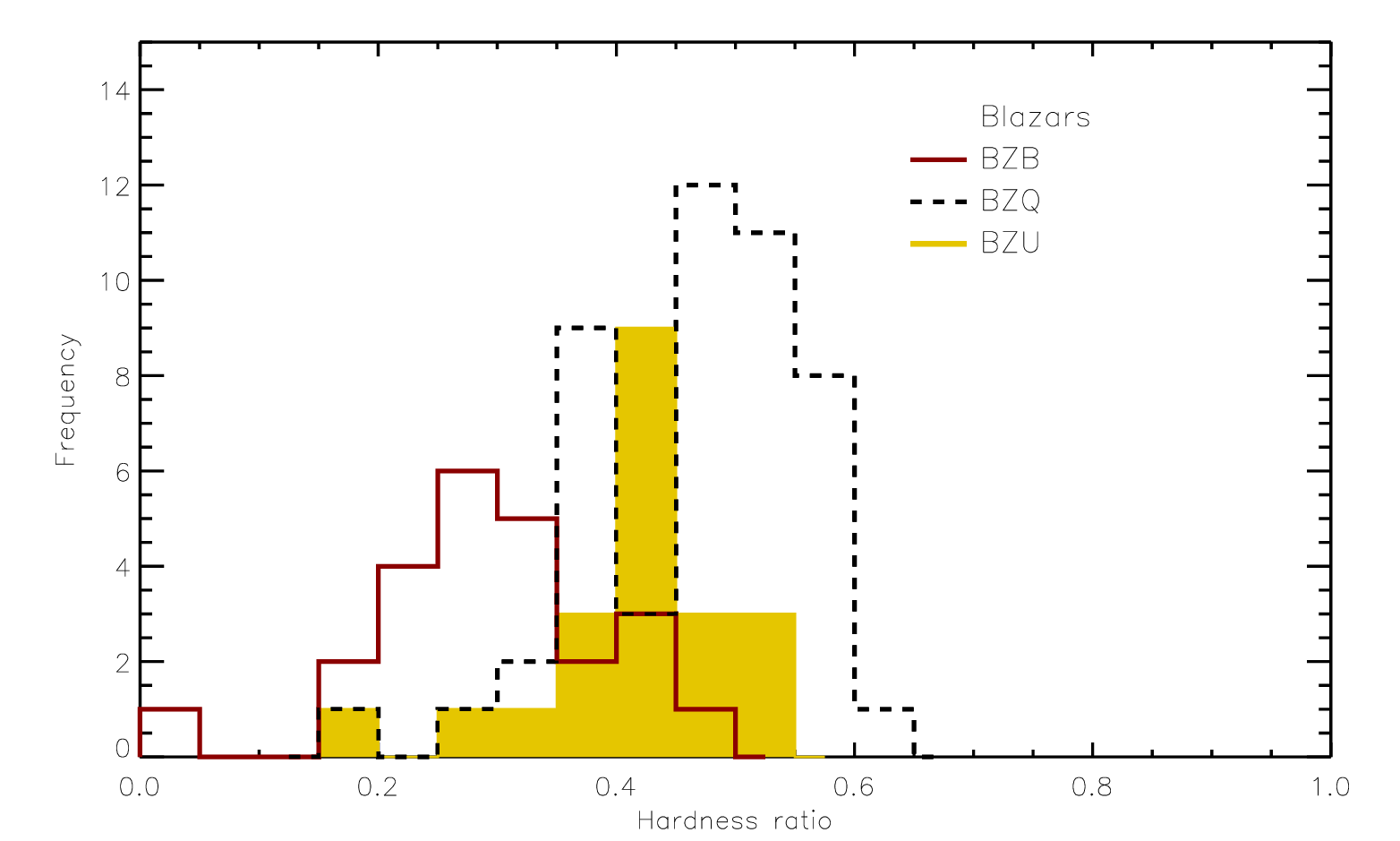,width=8.5cm,angle=0} }
\centerline{\psfig{figure=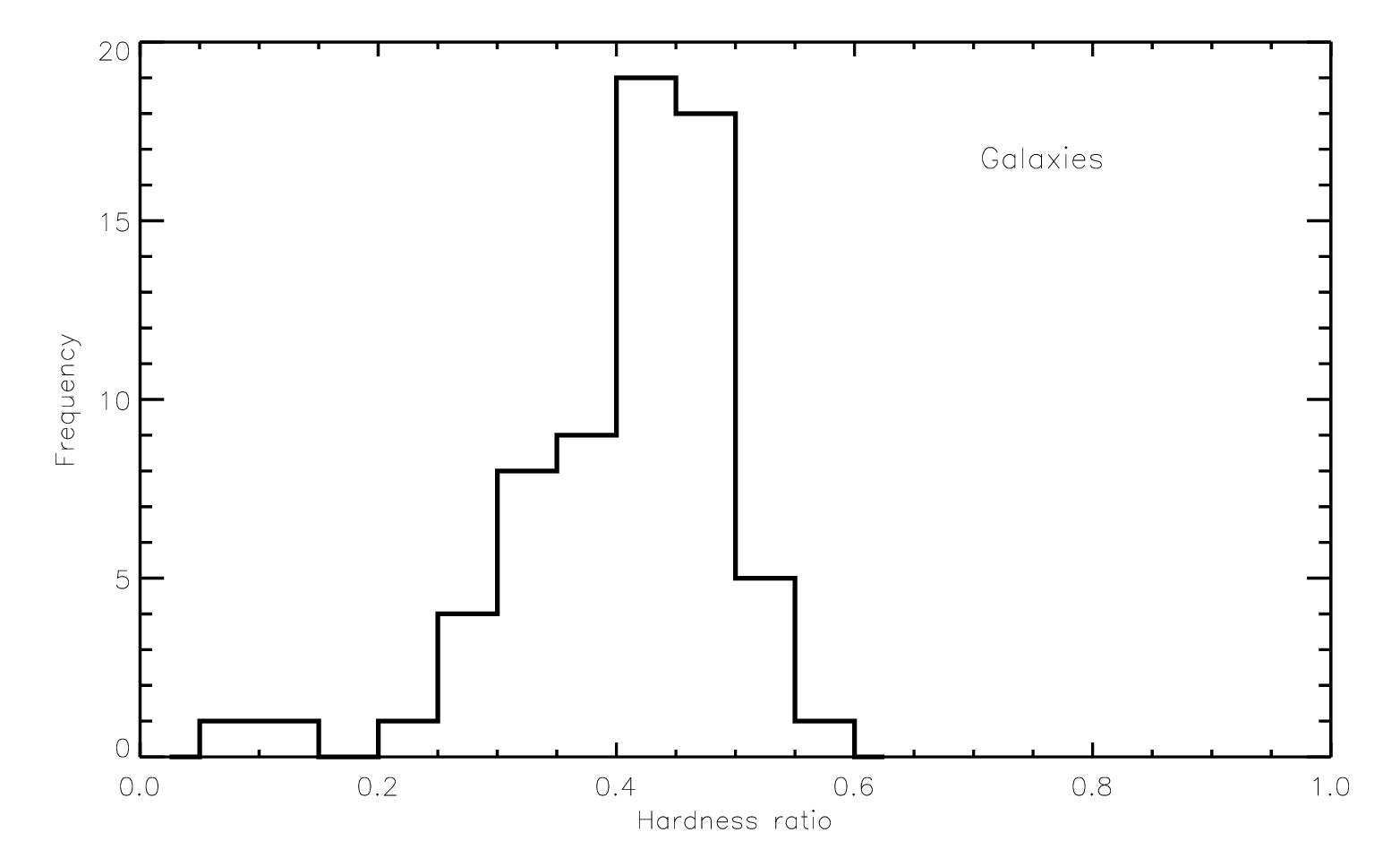,width=8.5cm,angle=0} \psfig{figure=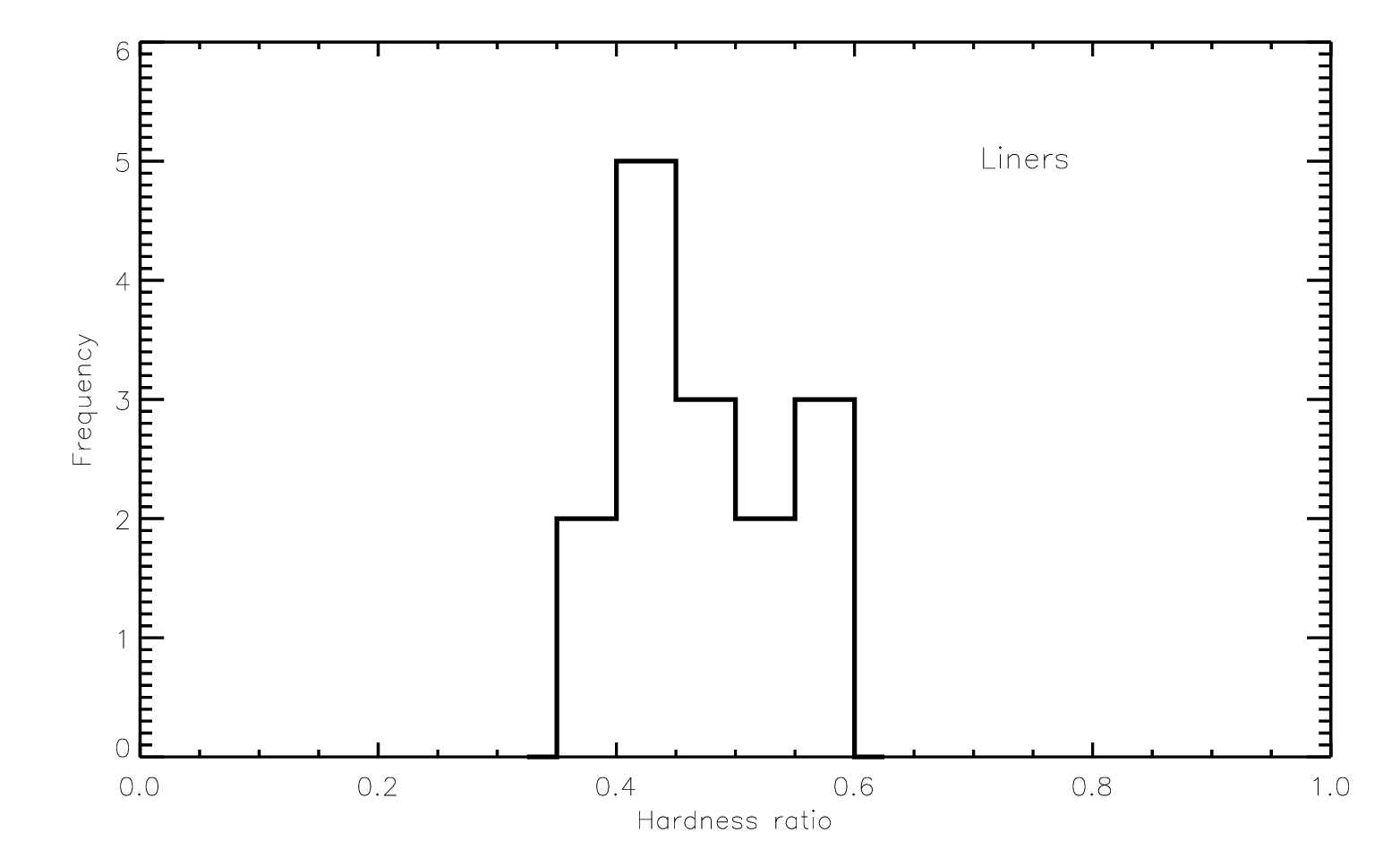,width=8.5cm,angle=0} }
\centerline{\psfig{figure=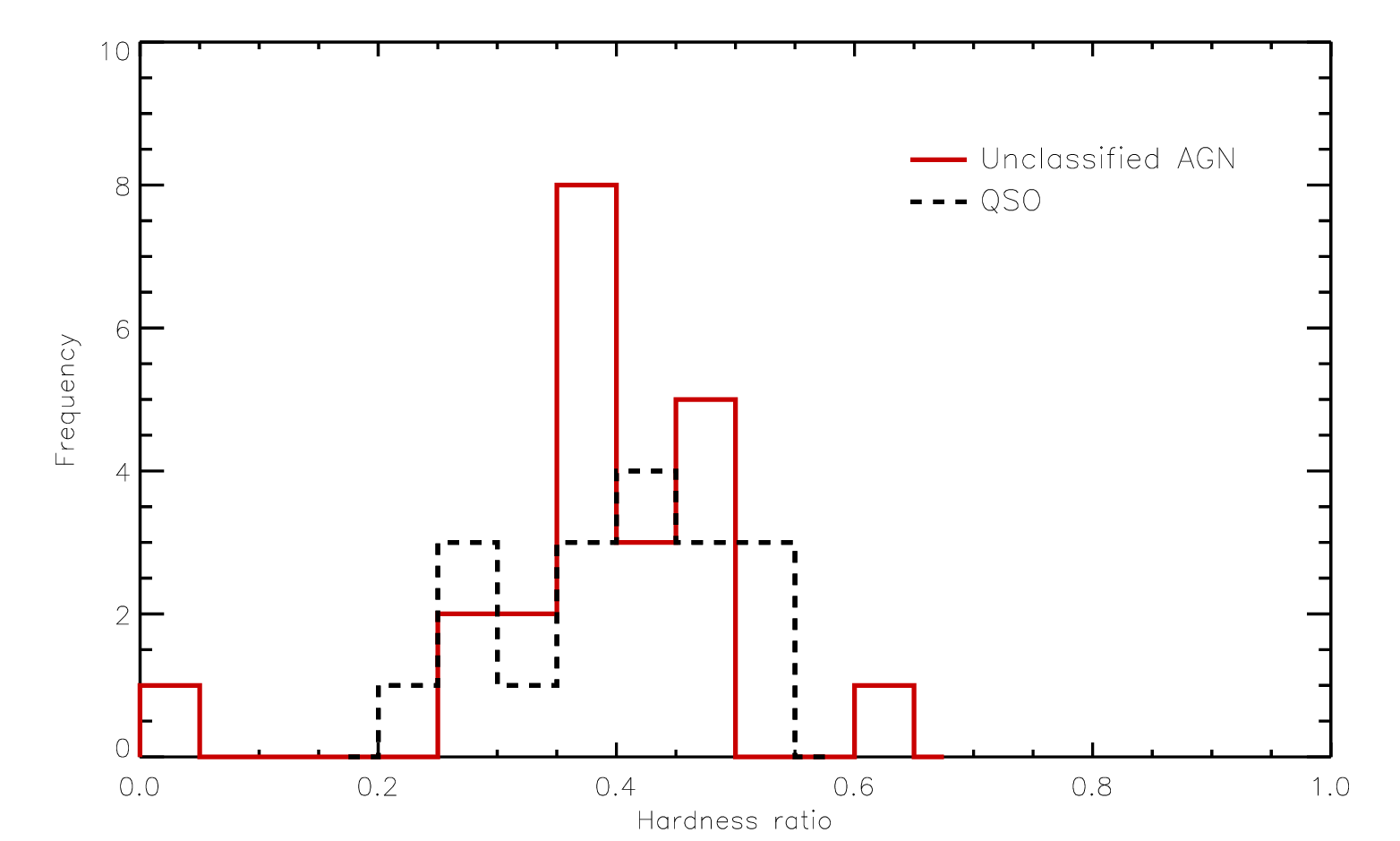,width=8.5cm,angle=0} \psfig{figure=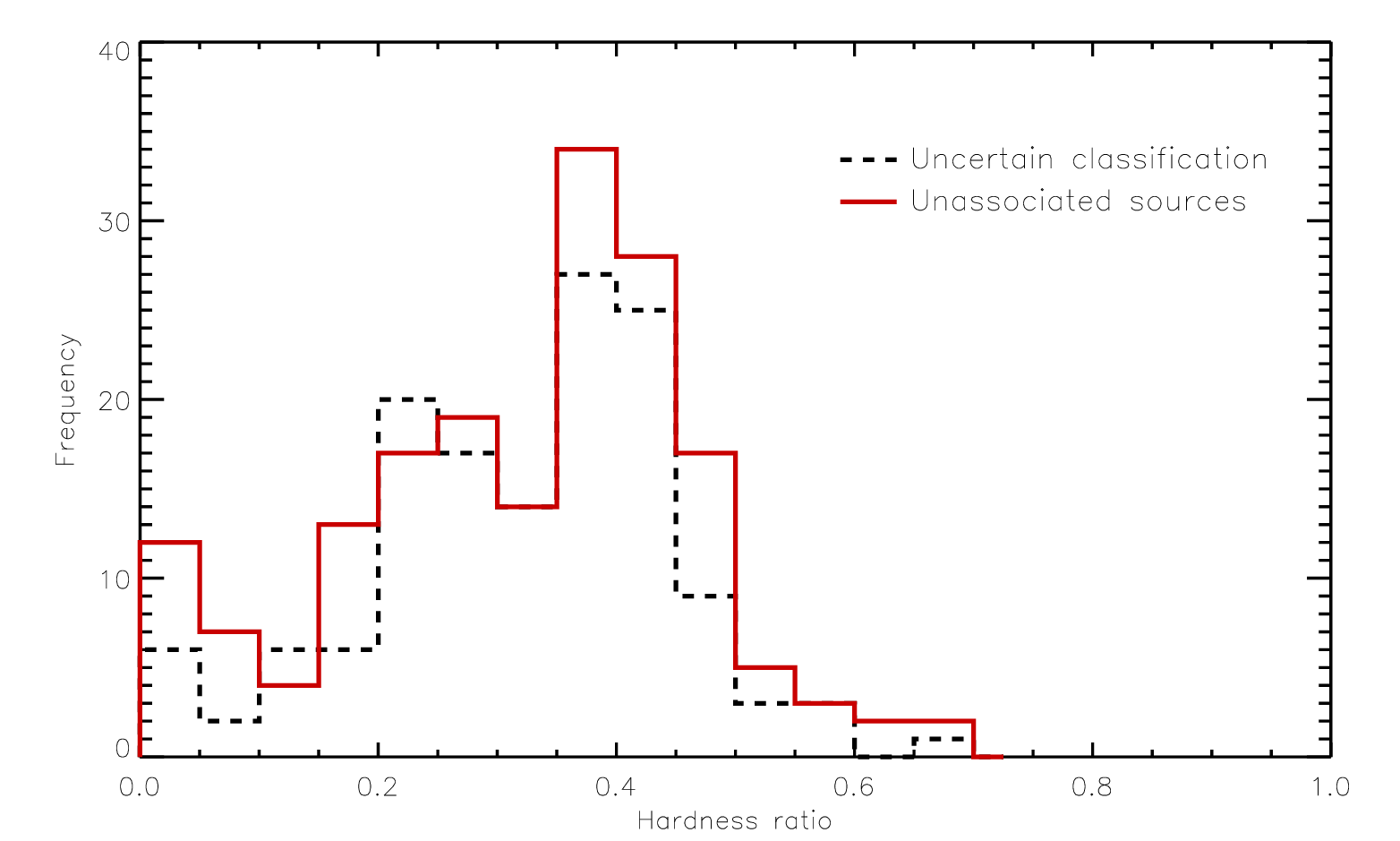,width=8.5cm,angle=0} }
\caption{Hardness ratio (35-150 keV)/(15-150 keV) distributions for the 
different classes of objects detected in the 54 months of BAT survey. \label{hr}}
\end{figure*}

Figure~\ref{hr} shows the HR distribution for each class of objects. As expected, the HR distribution 
for BZB is softer than for BZQ: this difference arises because the 15--150 keV band samples the high energy 
tail of the  synchrotron  peak for BZB and the rising part of the Compton peak for BZQ.
Blazars of 
uncertain classification (BZU) show an intermediate HR distribution. Clusters of galaxies fall in a very narrow
region of soft HR: we verified that their spectral distribution is consistent with the tail of a thermal emission
with kT $\sim 10$ keV, except for one object (CIZA~J0635.0+2231), with HR=0.26,
where we find evidence of hard non-thermal emission that may be related to the AGN content of the cluster.
The catalogue lists {\bf 67} objects classified as normal galaxies. The HR distribution of these 
sources peaks at $\sim0.4$ in a similar way to the other classes of active galaxies. This 
suggests that these objects may also contain an active nucleus.

The HR distribution of the sources with uncertain classifications and of unassociated sources suggests that most 
of these are of extragalactic nature.

\begin{figure}
\centerline{\psfig{figure=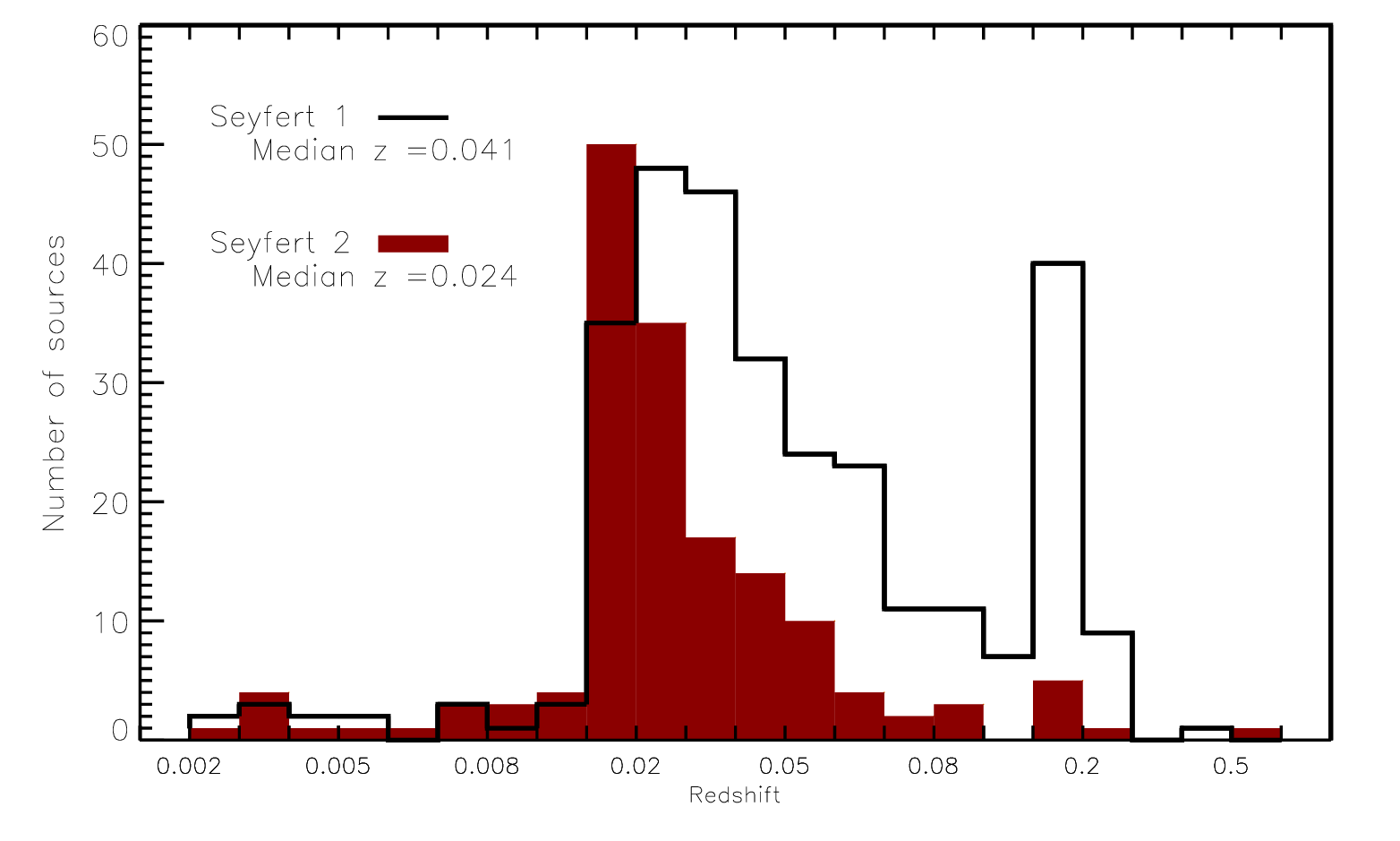,width=9cm,angle=0.}}
\centerline{\psfig{figure=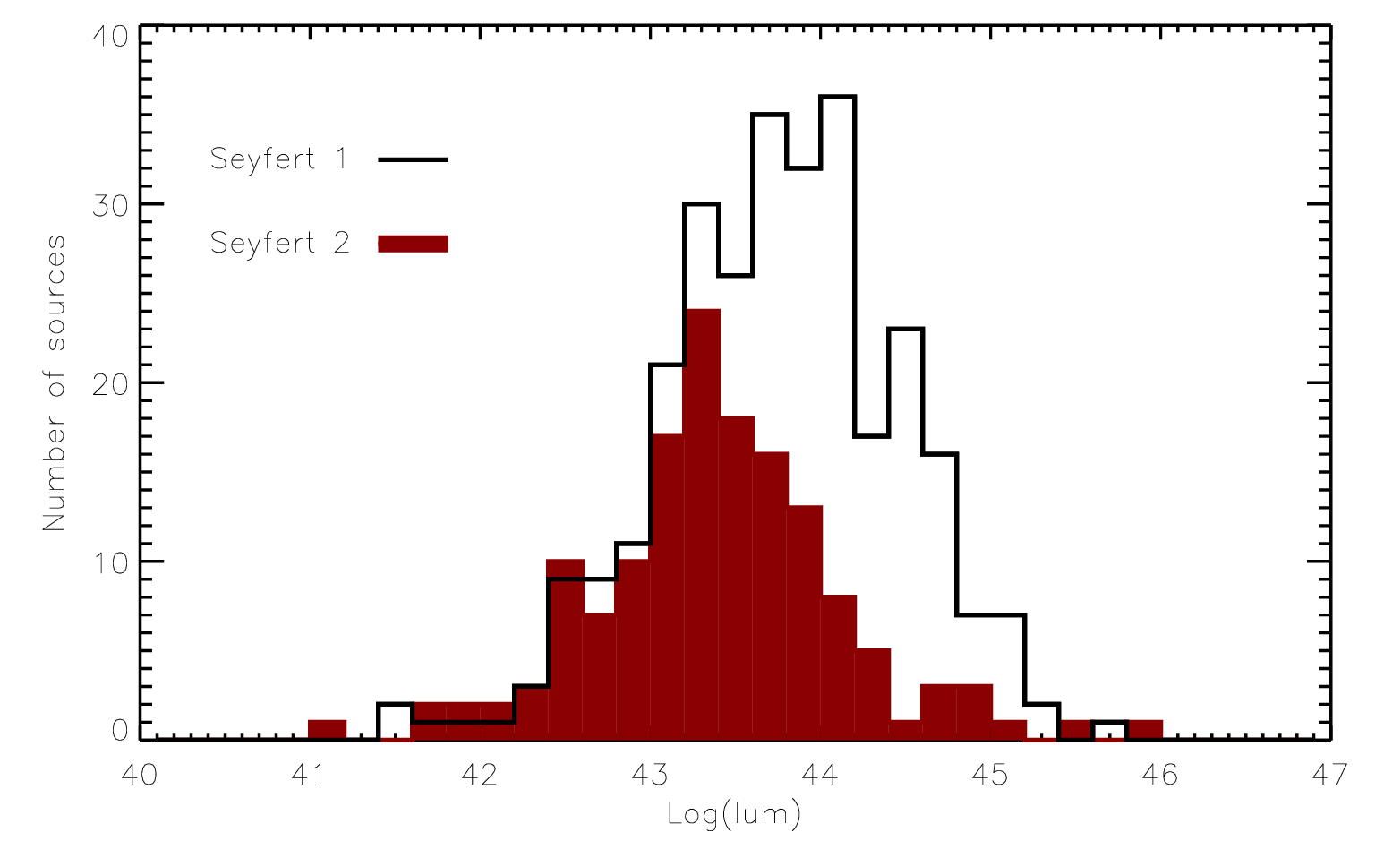,width=9cm,angle=0.}}
\caption{Redshift distribution (top) and 
luminosity distribution (bottom) of the Seyfert galaxies. \label{sylum}}

\end{figure}

Figure~\ref{sylum} shows the distribution of the redshift (top panel) and luminosity (bottom panel)
of the Seyfert 1 and Seyfert 2 galaxies included in the 54-month catalogue. The median of the 
redshift distribution is higher for Seyfert 1s (\~{z}$_{Sy1}$=0.040) than for Seyfert 2s 
(\~{z}$_{Sy2}$=0.025). The luminosity distribution shows that
Seyfert  1s are intrinsically more luminous than Seyfert 2s.

\subsection{The 54-month BAT catalogue and the INTEGRAL-ISGRI catalogue}

\begin{figure}
\centerline{\psfig{figure=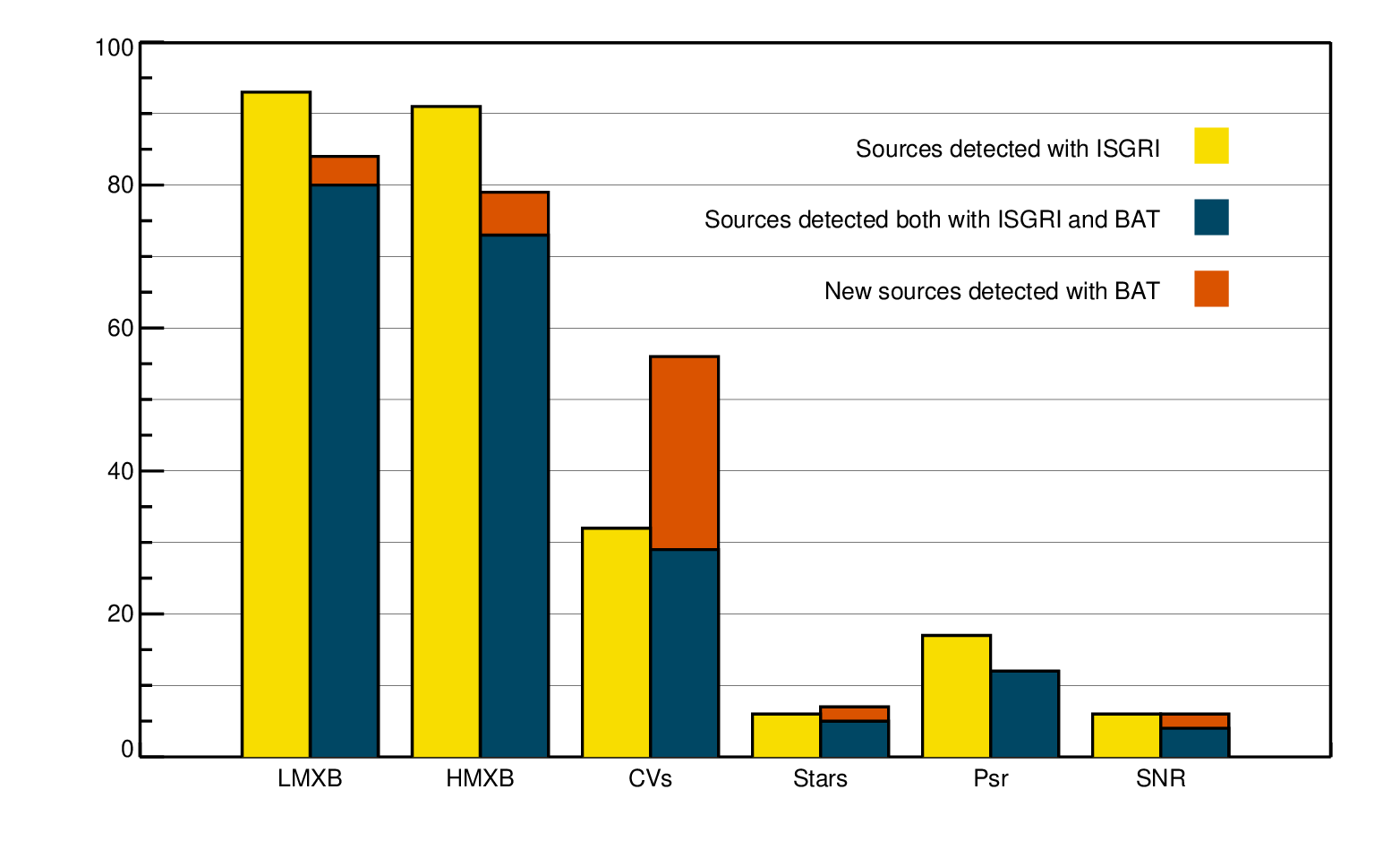,width=9cm,angle=0} }
\centerline{\psfig{figure=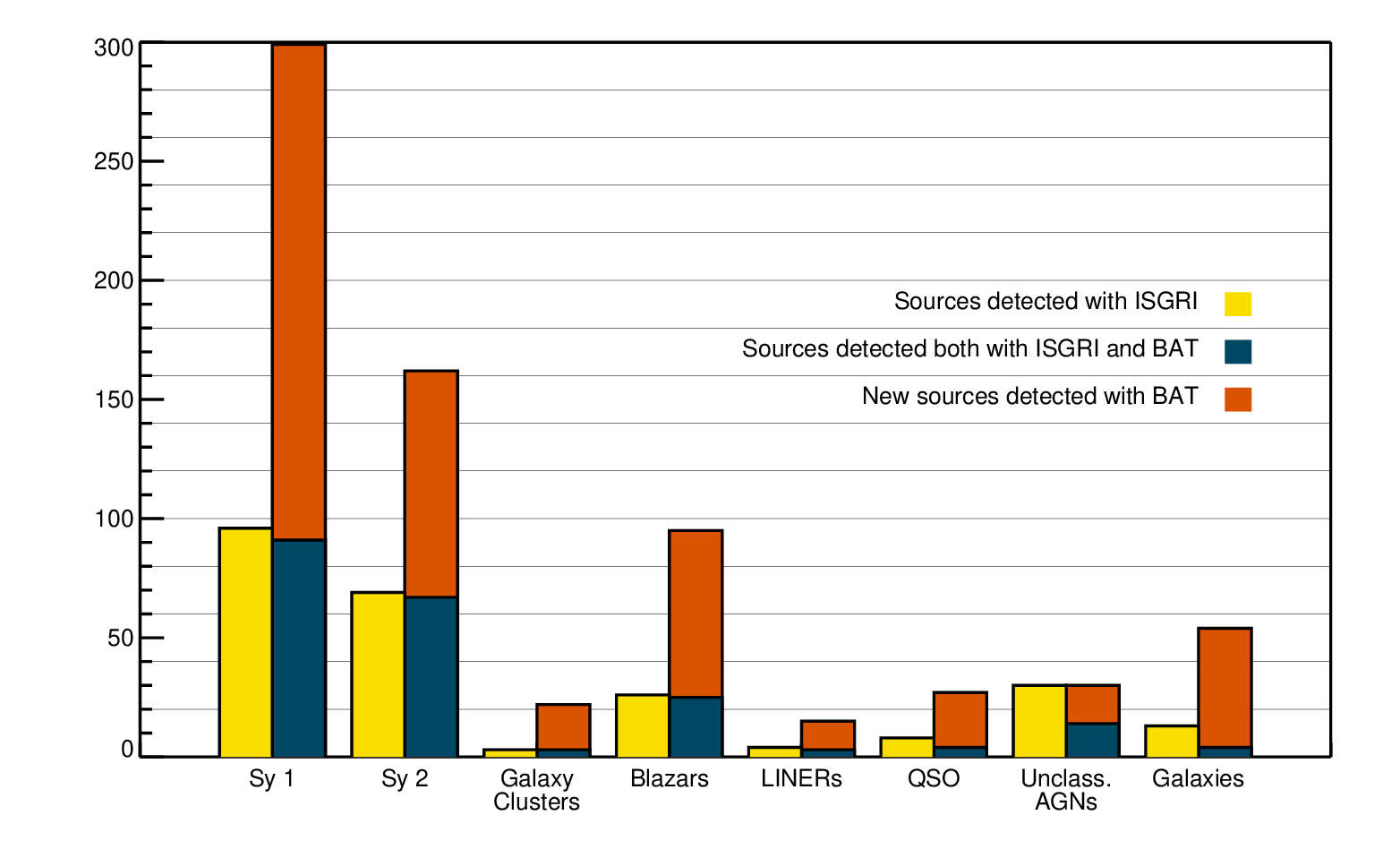,width=9cm,angle=0} }
\caption{Comparison between the sources in our catalogue and those
detected with ISGRI and reported in the INTEGRAL General Reference Catalogue V31. Top: Galactic sources.
Bottom: extragalactic sources.\label{integral}}
\end{figure}

We compared the sources detected in the 54-month BAT all-sky mosaic with those 
detected by INTEGRAL-ISGRI and reported in the INTEGRAL General Reference Catalogue 
V.31. 
The results are plotted in Fig.~\ref{integral}. 
For each object class, we report the sources detected by each of the two telescopes, 
highlighting those detected only by BAT.
While ISGRI dedicated most of the first years of its mission to a deep scan of the Galactic plane, 
BAT has taken advantage of its larger (with respect to ISGRI) field of view and different 
pointing strategy to achieve a uniform exposure of the whole sky. 
Within the Galactic sample, the number of low mass and high mass X-ray binaries 
is marginally higher in the ISGRI catalogue, although  10 sources are detected only with BAT. 
These sources have a transient behaviour and
are captured by BAT thanks to its larger field of view or because they are located in regions 
of low ISGRI exposure. BAT also detects a much larger sample (nearly a factor of 2) of 
cataclysmic variables, which are located mostly outside the Galactic plane. 
The BAT  extragalactic sample is a factor of between 2 and 3 larger than the ISGRI sample,
depending on the object class. This is
expected because of the lower limiting flux reached by BAT outside the Galactic plane.

\subsection{The 54-month BAT catalogue and the Fermi Large Area Telescope First 
Source Catalogue}

We compared our BAT catalogue with the Fermi Large Area Telescope First Source 
Catalogue \citep{abdo10} by searching for BAT sources whose position falls inside
the error box of each Fermi 
sources\footnote{http://fermi.gsfc.nasa.gov/ssc/data/access/lat/1yr\_catalog/}.
 
We found 59 BAT/Fermi correspondences to be associated with the same counterpart: 
16 BZBs, 27 BZQs, 5 BZUs, 3 Seyfert galaxies, 1 interacting galaxy, 3 high mass X-ray
binaries, and 4 pulsars/supernova remnants. 
Moreover, there are 4 BAT/Fermi correspondences with different counterpart association,
and 10 BAT/Fermi correspondences for which the Fermi source has not been associated with 
any counterpart. These 14 sources have been 
flagged with '?' in Col. 18 of Table~\ref{srctab}.

\begin{figure}
\centerline{\psfig{figure=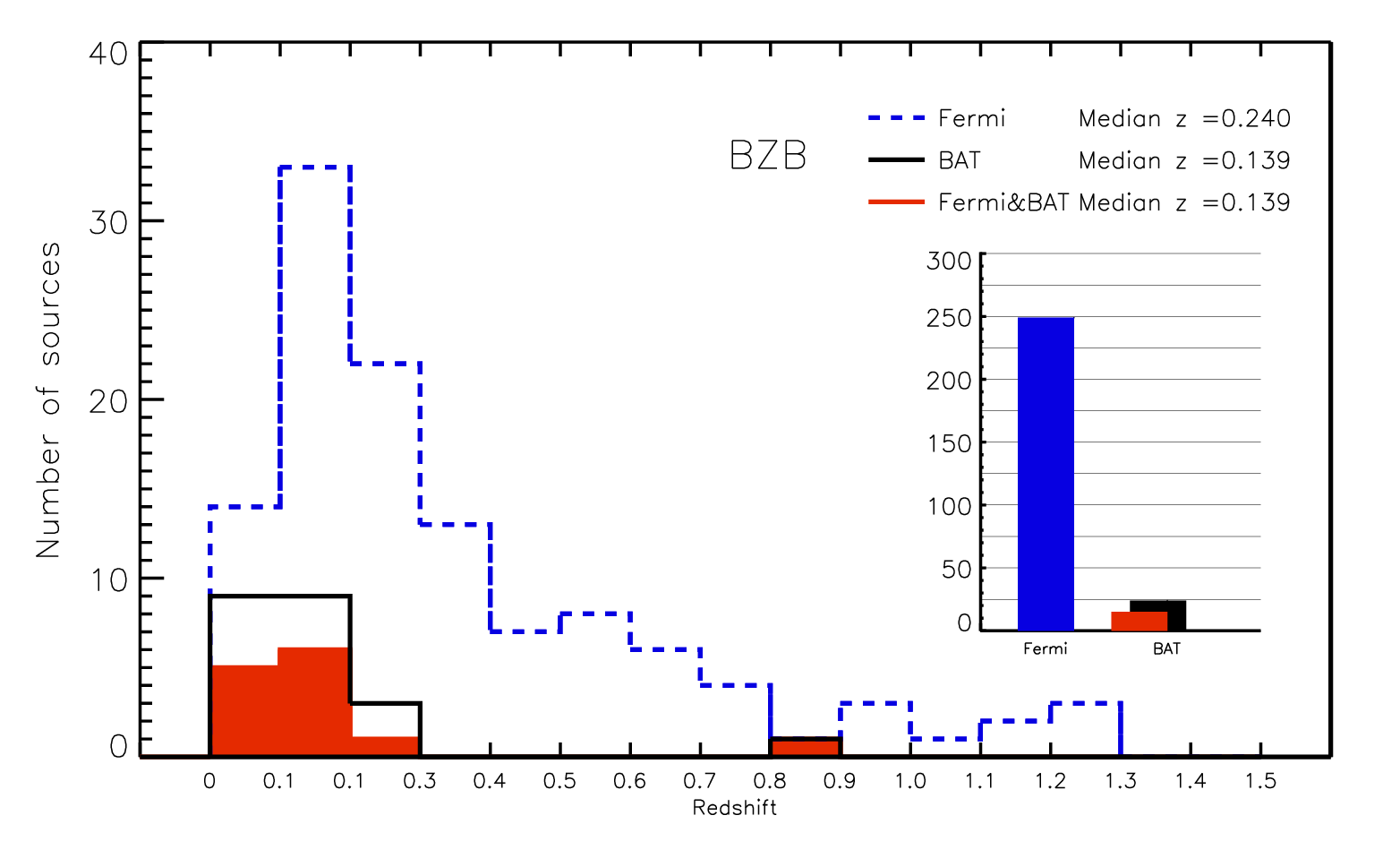,width=9cm,angle=0} }
\centerline{\psfig{figure=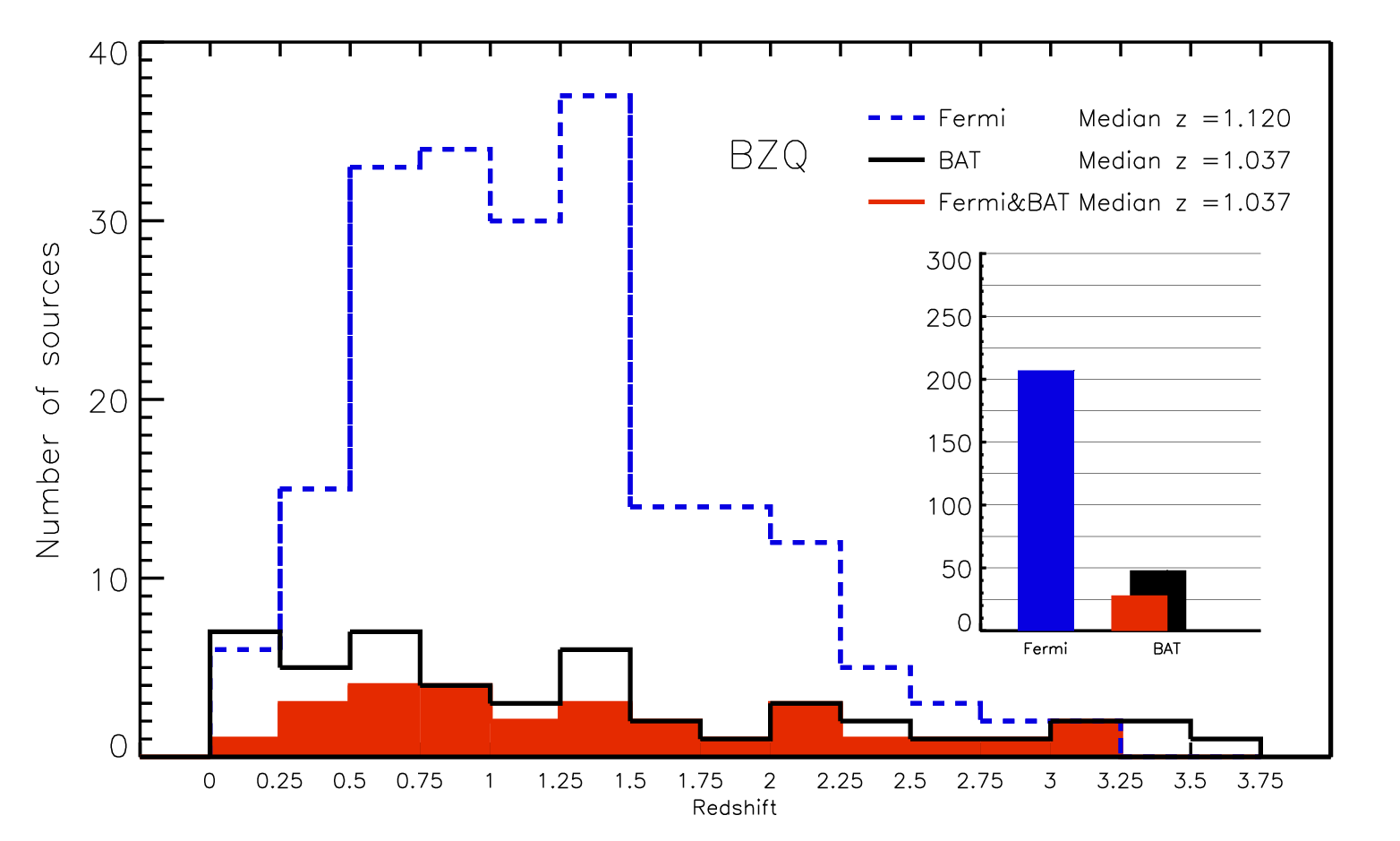,width=9cm,angle=0} }
\caption{Redshift distribution of the BZB (top) and  BZQ (bottom) sources.
The red line, the black dashed line, and the shaded blue area refer to blazars detected by
Fermi, {\it Swift}--BAT, and common to both catalogues, respectively. 
The insets show the total number of blazars in each sample.\label{blazars}}
\end{figure}

The largest sample of common sources is the blazar sample. In line with our association strategy,
we considered only Fermi blazars with a correspondence in the BZCAT. 
Figure~\ref{blazars} shows the redshift distributions of the selected common samples, 
superimposed on the redshift distributions of the whole Fermi and BAT blazar samples. 
The median of the redshift distribution for BZB is a factor of 2 higher for the Fermi sample 
than the BAT one, while the common sample has value in between these two. The median of the 
BAT and Fermi BZQ redshift distributions are very similar. The most distant blazar, 87GB 224928.1+22014
at z$\sim3.667$, is detected only by BAT.

\section{Conclusions\label{concl}}
We have analysed the BAT hard X-ray survey data of the first 54 months of the
{\it Swift} mission. The 5 $\sigma$ 15--150 keV survey flux limit achieved on 50\% of the sky is $\sim0.9 \times10^{-11}$
\ferg (0.43 mCrab). 

We have compiled all-sky maps for three energy bands (15--30 keV, 15--70 keV, and 15--150 keV) and searched for excesses above
a significance threshold of 4.8 standard deviations.
The final catalogue, obtained by cross-correlating and merging the lists of excesses detected 
in the three energy bands, contains 1256 source candidates.
For each of them, we have searched for counterparts at lower energies using two different strategies. 
First we have analysed archival soft X-ray observations covering the position of the BAT excesses, applying count rate thresholds to select the most
likely counterparts (strategy A). With this strategy, we have been able to associate 920 BAT
excesses with a single softer counterpart; for 8 BAT excesses, we found two possible
counterparts.
The BAT excesses lacking any association after strategy A were cross-correlated with  
a list of possible counterparts compiled by merging several source lists (X-ray binaries, 
cataclysmic variables, supernova remnants, pulsars, cluster of galaxies, different classes of 
active galaxies, already known soft X-ray and $\gamma$-ray sources).
This second strategy (strategy B) enabled us  to associate 151 BAT sources with counterparts 
(18 with a double association, 2 with a triple association).
The final catalogue contains 1079 BAT sources with at least one associated counterpart 
and 177 unassociated sources ($\sim14\%$). The latter will be the subject of a follow-up 
campaign with {\it Swift}-XRT in the immediate future.
The sources among the different object classes consist of $\sim19\%$ Galactic sources, 
$\sim57\%$ extragalactic sources, and
$\sim10\%$ sources with a counterpart at softer energies whose nature has not yet been determined.

The counterpart of 563 of the 1079 BAT sources with at least one associated counterpart is 
coincident with a bright ROSAT source, while 83 BAT sources have a counterpart consistent with the position  
of a faint ROSAT source. The remaining BAT counterparts (640) do not have any ROSAT correspondence. 
This may be the signature of either moderate or strong
absorption preventing detection in the ROSAT energy band. 

Compared to the INTEGRAL-IBIS telescope, BAT has detected a much larger number of 
extragalactic sources. This difference is mainly due to the different fields of 
view of the two telescopes and their different observing strategies.

The comparison of our BAT catalogue with the Fermi Large Area Telescope First Source Catalogue
\citep{abdo10} has established that 59 BAT/Fermi sources are associated with the same counterpart: 
16 BZBs, 27 BZQs, 5 BZUs, 3 Seyfert galaxies, 1 interacting galaxy, 3 high mass X-ray
binaries, and 4 pulsars/supernova remnants. These small number of correspondences clearly 
indicates that the sky at these two different energy ranges is 
populated by different source types.

\begin{acknowledgements}

This research has made use of NASA's Astrophysics Data System Bibliographic Services,  
of the SIMBAD database, operated at CDS, Strasbourg, France, as well as of the 
NASA/IPAC Extragalactic Database (NED), which is operated 
by the Jet Propulsion Laboratory, California Institute of Technology, under contract with 
the National Aeronautics and Space Administration. 
This work was supported by contract ASI/INAF
I/011/07/0.

\end{acknowledgements}

\bibliographystyle{aa}
{}

\clearpage
\scriptsize
\onecolumn
\begin{landscape}
\begin{longtable}{rlllrrrrrlrrrrll}
\caption{BAT 54-month catalogue\label{srctab}}\\
\hline\hline
   & PBC name & ID & Type$^a$ & RA  & Dec & Error radius& Offset & SNR & Flux$^b$& Hardness ratio      &Redshift&$\log{L_{14-150}}^c$ &Var &Flag 1$^d$&Flag 2$^e$\\
   &            &    &      &(deg)&(deg)&(arcmin)  &(arcmin)&   &(\ferg)&$(R_{30-150}/R_{14-150})$&        &(erg s$^{-1})$ &&A B C D &I R F \\ 
\hline
\endfirsthead
\caption{continued.}\\
\hline\hline
   & PBC name & ID & Type$^a$ & RA  & Dec & Error radius& Offset & SNR & Flux$^b$& Hardness ratio      &Redshift&$\log{L_{14-150}}^c$ &Var &Flag 1$^d$&Flag 2$^e$\\
   &            &    &      &(deg)&(deg)&(arcmin)  &(arcmin)&   &(\ferg)&$(R_{30-150}/R_{14-150})$&        &(erg s$^{-1})$ &&A B C D &I R F \\ 
\hline
\endhead
\hline
\endfoot
    \bf  1 & \bf 2PBC J0000.3+0416&                      ...& ... &   0.092&   4.270 &4.99&     ...&    5.03&$   0.7\pm0.5$&$       0.4\pm0.2 $&   ...&      ...&      1.34&1~~N~~h~~ & ~~~~~~~\\
    \bf  2 & \bf 2PBC J0000.9-0708&                      ...& ... &   0.234&  -7.134 &4.42&     ...&    5.58&$   1.4\pm0.7$&$     0.33\pm0.09 $&   ...&      ...&      1.59&2~~N~~h~~ & ~~~~~~~\\
    \bf  3 & \bf 2PBC J0001.8-7658&                      ...& ... &   0.466& -76.982 &4.46&     ...&    5.50&$   0.9\pm0.6$&$       0.6\pm0.2 $&   ...&      ...&      1.91&3~~N~~h~~ & ~~~~~~~\\
    \bf  4 & \bf 2PBC J0002.5+0320&                 NGC 7811& Sy1 &   0.643&   3.337 &4.61&    2.12&    5.20&$   0.9\pm0.6$&$       0.5\pm0.2 $&0.0255&    43.12&      1.44&3~~N~~h~~b& ~~~b~~~\\
    \bf  5 & \bf 2PBC J0004.1+7017&  2MASX J00040192+7019185& AG? &   1.035&  70.298 &4.90&    1.54&    5.17&$   0.9\pm0.6$&$       0.4\pm0.1 $&0.0960&    44.32&      1.38&3~~N~~h~~a& I~~~~~~\\
         6 &     2PBC J0006.3+2012&                  Mrk 335& Sy1 &   1.591&  20.199 &2.78&    0.60&   12.55&$   1.4\pm0.8$&$     0.25\pm0.06 $&0.0254&    43.33&      1.80&3~~Y~~h~~a& ~~~b~~~\\
    \bf  7 & \bf 2PBC J0009.3-0034&  2MASX J00091156-0036551& Sy2 &   2.338&  -0.576 &4.02&    3.35&    6.56&$   0.8\pm0.5$&$       0.4\pm0.1 $&0.0732&    44.03&      1.61&1~~N~~h~~b& ~~~~~~~\\
           &                      & [VV2003c] J000939.5-0032& Sy1 &        &         &    &    5.06&        &$            $&$                 $&0.0400&    43.48&          &          &        \\
\end{longtable}
\begin{list}{}{}
    \item[$^a$] The source type is coded according to the nomenclature used in
    SIMBAD, except for the blazars included in the Roma-BZCAT \citep{massaro09}, for which we 
    adopted the relevant nomenclature.
    \item[$^b$] Flux is in units of $10^{-11}$\ferg.
    \item[$^c$] In case of more than one counterpart, the luminosity is calculated for each
    counterpart (where a redshift is available) assuming that it produces all the observed flux.
    \item[$^d$] Flag A: energy band with highest significance (1=15--150
    keV; 2=15--30 keV; 3= 15--70 keV);\\
    Flag B: "Y" if already reported as hard X-ray source;\\
    Flag C: "l" if the source has $|b|<5^{\circ}$, "h" if the source has $|b|>5^{\circ}$;\\
    Flag D: strategy used for the identification (see Sect.\ref{id}).
    \item[$e$] 
    Flag I: "I" if source seen by INTEGRAL\\
    Flag R: "b" if correlated with a ROSAT Bright source, "f" if correlated with a
    ROSAT Faint source;\\
    Flag F: "F" if the counterpart is associated to a Fermi source; "?" if the BAT
    position is consistent with the Fermi position but the associate counterparts
    are different.
\end{list}
\end{landscape}
\clearpage

\end{document}